\documentclass[12pt,preprint]{aastex}

\slugcomment{CDM Distribution in the Eddington Theory.}

\shorttitle{New CDM Velocity Distribution}
\shortauthors{Vergados and Owen}

\newcommand{\newc}{\newcommand} 
\newc{\lra}{\leftrightarrow} 
\newc{\beq}{\begin{equation}} 
\newc{\eeq}{\end{equation}} 
\newc{\barr}{\begin{eqnarray}} 
\newc{\earr}{\end{eqnarray}} 
\begin{document}

%% LaTeX will automatically break titles if they run longer than
%% one line. However, you may use \\ to force a line break if
%% you desire.

\title{New Velocity Distribution for Cold Dark Matter in the Context of the
Eddington Theory}

%% Use \author, \affil, and the \and command to format
%% author and affiliation information.
%% Note that \email has replaced the old \authoremail command
%% from AASTeX v4.0. You can use \email to mark an email address
%% anywhere in the paper, not just in the front matter.
%% As in the title, you can use \\ to force line breaks.

\author{J.D. Vergados\altaffilmark{1}}
\affil {Department of Physics, Unisa, Pretoria, South Africa.}
\email{vergados@cc.uoi.gr}
\and
\author{D. Owen \altaffilmark{2}}
\affil{Department of Physics, Ben Gurion University, Israel.}
\email{owen@bgumail.bgu.ac.il}
%% Notice that each of these authors has alternate affiliations, which
%% are identified by the \altaffilmark after each name.  Specify alternate
%% affiliation information with \altaffiltext, with one command per each
%% affiliation.

\altaffiltext{1}{Permanent address:Theoretical Physics Division,
 University of Ioannina, Ioannina, Gr 451 10, Greece.}
\altaffiltext{2}{Visiting the Theoretical Physics Division,
 University of Ioannina, Ioannina, Gr 451 10, Greece.}

%% Mark off your abstract in the ``abstract'' environment. In the manuscript
%% style, abstract will output a Received/Accepted line after the
%% title and affiliation information. No date will appear since the author
%% does not have this information. The dates will be filled in by the
%% editorial office after submission.

\begin{abstract}
Exotic dark matter together with the vacuum energy (associated with the 
cosmological constant) seem to dominate the Universe.
Thus its direct detection 
is central to particle physics and cosmology. Supersymmetry
provides a natural dark matter candidate, the lightest
supersymmetric particle (LSP). One essential ingredient in obtaining the direct
detection rates is the density and velocity distribution of the LSP. The 
detection rate is proportional to this density in our vicinity. Furthermore,
since this rate is expected to be very low, one should explore
the two characteristic signatures of the process, namely
 the modulation effect, i.e. the dependence of
the event rate on  the Earth's motion and the correlation of the directional
rate with the motion of the sun. Both of these crucially depend on the LSP
velocity distribution. In the present paper we study simultaneously density
profiles and velocity distributions based on the Eddington theory.
\end{abstract}

%% Keywords should appear after the \end{abstract} command. The uncommented
%% example has been keyed in ApJ style. See the instructions to authors
%% for the journal to which you are submitting your paper to determine
%% what keyword punctuation is appropriate.
\keywords{Cosmology:Eddington theory, velocity profiles, rotational curves-
Cold Dark Matter:velocity distribution, direct detection rates.}
%%%%%%%%%%%%%%%%%%%%%%%%%%%%%%%%%%%%%%%%%%%%%%%%%%%%%%%%%%%%%%%%%%%%%
%%%%%%%%%%%%%%%%%%%%%%%%%%%%%%%%%%%%%%%%%%%%%%%%%%%%%%%%%%%%%%%%%%%%%
\date{\today}
\newpage
%%%%%%%%%%%%%%%%%%%%%%%%%%%%%%%%%%%%%%%%%%%%%%%%%%%%%%%%%%%%%%%%%%%%%
\section{Introduction}
In recent years the consideration of exotic dark matter has become necessary
in order to close the Universe \citep{Jungm}.
 The COBE data ~\citep{COBE} suggest that CDM (Cold Dark Matter)
component is at least is at least $60\%$ ~\citep{GAW,Gross} of the total mass.
 On the other hand evidence from two different teams,
the High-z Supernova Search Team \citep{HSST} and the
Supernova Cosmology Project  ~\citep{SPF} $^,$~\citep{SCP1,SCP2} 
  suggests that the Universe may be dominated by 
the  cosmological constant $\Lambda$.
Thus the situation can be adequately
described by  a baryonic component $\Omega_B=0.1$ along with the exotic 
components $\Omega _{CDM}= 0.3$ and $\Omega _{\Lambda}= 0.6$.
In a more detailed $\Lambda CDM$ analysis ~\citep{Primack} one finds:
$$\Omega_b=0.040\pm0.002,~\Omega_m=\Omega_{CDM}=0.33\pm0.035,~\Omega_{HDM}\le0.05,~\Omega_{\Lambda}=0.73\pm0.08$$
(see also \citep {Turner} and \citep{Einasto}).
%gives $\Omega_{m}= \Omega _{CDM}+ \Omega _B=0.4$.
Since the non exotic component cannot exceed $40\%$ of the CDM 
~\citep{Jungm},~\citep {Alcock}, there is room for the exotic WIMP's 
(Weakly  Interacting Massive Particles).
  In fact the DAMA experiment ~\citep {BERNA2} 
has claimed the observation of one signal in direct detection of a WIMP, which
with better statistics has subsequently been interpreted as a modulation signal
~\citep{BERNA1,BERNA3}.

 In the most favored scenario of supersymmetry the
LSP can be simply described as a Majorana fermion, a linear 
combination of the neutral components of the gauginos and Higgsinos
\citep{Jungm,ref1,Gomez1,Gomez2,Gomez3,gtalk,ref2a,ref2b,ref2c,ref2d}. 

 Since this particle is expected to be very massive, $m_{\chi} \geq 30 GeV$, and
extremely non relativistic with average kinetic energy $T \leq 100 KeV$,
it can be directly detected ~\citep{JDV96,Spira,KVprd} mainly via the recoiling
of a nucleus (A,Z) in elastic scattering.

In order to compute the event rate one needs the following ingredients:

1) An effective Lagrangian at the elementary particle 
(quark) level obtained in the framework of supersymmetry as described 
, e.g., in ~\citep{Jungm,ref2a,ref2b,ref2c,ref2d}. 

2) A procedure in going from the quark to the nucleon level, i.e. a quark 
model for the nucleon. The results depend crucially on the content of the
nucleon in quarks other than u and d. This is particularly true for the scalar
couplings as well as the isoscalar axial coupling
 ~\citep{Dree1,Dree2,Chen1,Chen2}.

3) Compute the relevant nuclear matrix elements \citep{Ress,DIVA00}
using as reliable as possible many body nuclear wave functions.
The situation is a bit simpler in the case of the scalar coupling, in which
case one only needs the nuclear form factor.

4) The LSP density and velocity distribution. Among other other things, since
 the detection rates are expected to be very small, the velocity distribution
is crucial in exploiting the characteristic experimental signatures provided by
the reaction, namely: a) the modulation of the event rates due to the earth's
revolution around the sun \citep{Verg98,Verg99}$^-$\citep{Verg00} and b) the
 correlation of the rates with the Sun's direction of motion in directional
 experiments, i.e. experiments in
which the direction of the recoiling nucleus is observed \citep {ref1,UKDMC}.
 To obtain the right density and velocity distributions is the purpose of the
present paper.

In the past various velocity distributions have been considered. The most 
popular one is the isothermal Maxwell-Boltzmann velocity distribution 
with $<\upsilon ^2>=(3/2)\upsilon_0^2$ where $\upsilon_0$ is the velocity of
the sun around the galaxy, i.e. $220~km/s$. Extensions of this M-B
distribution were also considered, in particular those that were axially
symmetric with enhanced dispersion in the galactocentric direction
 \citep {Druk,Verg00}. In such
distributions an upper cutoff $\upsilon_{esc}=2.84\upsilon_0$ was introduced
by hand.
 
Non isothermal models have also been considered. Among those one should
mention the late infall of dark matter into the galaxy, i.e caustic rings
 \citep{SIKIVI1,SIKIVI2,Verg01,Green,Gelmini}, and dark matter orbiting the
 Sun \citep{Krauss}.

The correct approach in our view is to consider the Eddington approach 
\citep{EDDIN}, i.e. to 
obtain both the density and the velocity distribution
from a mass distribution, which depends both on the velocity and
the gravitational potential. This approach has been extensively studied
by Merritt \citep{MERRITT} and recently applied to dark matter by Ullio and
Kamionkowski\citep{ULLIO}
 
\section{Density Profiles}
As we have seen in the introduction the matter distribution can be given
 as follows
\beq
dM=2\pi~f(\Phi(r),\upsilon_r,\upsilon_t)~dx~dy~dz
~\upsilon_t~d\upsilon_t~d\upsilon_r
\label{distr.1}
\eeq
where the function f the distribution function, which depends on $r$ through 
the potential $\Phi(r)$ and the tangential and radial velocities $\upsilon_t$
and $\upsilon_r$ (we assume axial symmetry in velocity space). Thus the density 
of matter $\rho$ satisfies the equation 
\beq
d\rho=2\pi~f(\Phi(r),\upsilon_r,\upsilon_t)~~\upsilon_t~d\upsilon_t~d\upsilon_r
\label{distr.2}
\eeq
It is more convenient instead of the velocities to use the total energy $E$ and the angular momentum $J$ via the equations
\beq
J=\upsilon_t~r~~,~~2E=\upsilon_r^2+\frac{J^2}{r^2}+2~\Phi(r)
\label{distr.3}
\eeq
The use of these variables, which are constants of motion, is very useful, when
one wants to study equilibrium states. We thus find
\beq
%\rho=\frac{2\pi}{r^2\upsilon^2}~\int~ f(E,J)~J}~dJ~dE
\rho=\frac{2\pi}{r^2}~\int~\int \frac{f(E,J)~J}{\sqrt{2(E-\Phi(r))-J^2/r^2}}
                                  ~dJ~dE
\label{distr.4}
\eeq
The limits of integration for $E$ are from $\Phi$ to $0$ and for $J$ from
$0$ to $[2r^2(E-\Phi(r))]^{1/2}$.

 Following Eddington we will choose a distribution function of the form
\beq
f(E,J)=K_{\lambda} (-2E)^{\lambda}
\label{distr.5}
\eeq
($E$ is negative for a bound system), where $\lambda$
is a parameter, which will depend on the type of matter, and $K_{\lambda}$ is
a normalization constant, which will be related to the density at some point.
 With this choice of the
distribution function it is quite straightforward to find the relationship
between the density $\rho$ and the potential. The result is
\beq
\rho=K_{\lambda}
      2^{\lambda+3/2} \pi |\Phi(r)|^{\lambda+3/2}~\beta(\lambda+1,3/2)
\label{distr.6}
\eeq
with
\beq
\beta(a,b)=\frac{\Gamma(a)~\Gamma(b)}{\Gamma(a+b)}
\label{distr.7}
\eeq
Eq. (\ref{distr.6}) can be cast in the simple form:
\beq
\rho = \rho_{\lambda}(0) (\chi(x))^{\lambda+3/2}
\label{distr.6b}
\eeq
with
\beq
\chi(x)=\frac{\Phi(x)}{\Phi_0},~~~x=r/r_s
\label{distr.6c}
\eeq
with $\Phi_0=\Phi(0)$ and $r_s$ the galactic radius (position of the sun).
The constant $K_\lambda$ is related to the density at the origin via the
relation:
\beq
K_{\lambda}=\frac{\rho_{\lambda}(0)}{\pi(2 |\Phi_0|)^{\lambda+3/2}
                   \beta(\lambda+1,3/2)}
\label{distrb.6d}
\eeq
The above distribution function is isotropic. Following Merritt 
\citep{MERRITT} we can
 introduce an anisotropy by modifying the distribution function as follows:
\beq
f(E,J)=K_{\lambda} (-2E)^{\lambda}(1\pm\frac{J^2}{r_a^2})
\label{distr.8}
\eeq
Instead of the parameter $r_a$ we find it convenient for our
 applications later on (see below) to  adopt the more recent conventions
and write the above equation as follows:
\beq
f(E,J)=K_{\lambda} (-2E)^{\lambda}[1+\alpha_s\frac{J^2}{(r_s\upsilon_m)^2}]
\label{distr.9}
\eeq
 where $\upsilon_m$ is the maximum velocity allowed by the potential, to be 
specified below,  and $\alpha_s$ the asymmetry parameter.
Proceeding as above we find that this induces a correction to the density of the
form:
\beq
\Delta\rho=K_{\lambda} \frac{4}{3} 2^{\lambda+3/2} \pi|\Phi(r)|^{\lambda+5/2}
           ~\beta(\lambda+1,5/2) \frac{\alpha_s}{\upsilon_m^2} x^2
\label{distr.10}
\eeq
%The above equations hold for both ordinary matter ($\lambda=7/2$) and
%dark matter ($\lambda=1/2$).

Combining Eqs (\ref{distr.6}) and (\ref{distr.10}) we get
\beq
\rho_{\lambda} (r) = \rho_{\lambda}(0) \psi_{\lambda}(x)
\label{dens.1a}
\eeq
\beq
\psi_{\lambda}(x) =(\chi(x))^{\lambda+3/2}
      [1+\frac{4}{3}~a~ \frac{\beta(\lambda+1,5/2)} {\beta(\lambda+1,3/2)} x^2
           \chi(x)]
\label{dens.1b}
\eeq
with $a=\alpha_s |\Phi_0|/\upsilon_m^2$.

 Since the scale of the potential appears 
only via the parameter $a$ one, in principle, could have two
$a$ parameters, one for Matter ($a_m$) and one for dark matter ($a_{dm}$). In 
the present work we will assume that they are equal.We remind the reader that
 $\alpha_s$ is the asymmetry parameter to be treated phenomenologically.

\section{Allowed Density Functions}
The central question is to specify 
 density $\psi_{\lambda}(x)$, the potential $\chi(x)$
 and the mass density distribution entering Eq. (\ref{distr.1}).
 One can adopt one of two procedures:

1) Start out with a given density, obtained, e.g., phenomenologically, and find
the potential by solving Poisson's equation. This way one obtains, at least 
parametrically, e.g.  with the radial coordinate as a parameter, the proper 
relation between the density and the potential. This approach has been adopted
by a lot of researchers, see e.g. \citep{WID}, \citep{EVAN}, \citep{HEN},
\citep{ULLIO}.
 From this one can obtain the mass density function $f(\Phi,\upsilon)$.
 This approach leads, in general, to non analytic, i.e.
complicated, velocity distribution, hard to implement in dark matter
calculations \citep{ULLIO}, especially for not spherically symmetric mass
distributions.

2)  Use the above simple relation between the density and the potential and solve
the differential equation resulting from Poisson's equation and thus obtain each
one of them. One hopes that this way one will get a density distribution, which 
describes adequately dark matter distribution globally, via fitting the rotational
curves, and in our vicinity. This approach can easily deal with asymmetric density
distributions. The simplicity of the relation between the density and the potential
thus yields the bonus that the obtained velocity distribution is analytic and can
 easily
implemented in obtaining the event rates for direct dark matter detections.It can
 also be extended to include more than one power in $\lambda$. In this paper we
will follow this approach, and leave it for the future to use semi-analytic
mass distributions obtained in the approach outlined in the previous paragraph.

For a spherically symmetric potential Poisson's equation leads to a 
differential equation of the type:
\beq
x \chi^{''}(x)+2 \chi^{'}(x)=-\Lambda~x~(\chi(x))^n
                            [1+ \frac{2a}{n+1} x^2 \chi(x)]
\label{difeq.1}
\eeq
with $n=\lambda+1/2$ and
\beq
\Lambda=-\frac{4 \pi G_N r^2_s \rho_{\lambda}(0)}{\Phi_0}
\label{def.1}
\eeq
with $G_N$ Newton's gravitational constant. The dimensionless quantity $\Lambda$
is assumed to be positive (attractive potential). Since the asymmetry parameter
is assumed to be small, the first term in the right hand side dominates.
 Introducing the variable $\xi=\sqrt{\Lambda} x$ and $\chi(x)=u(x \sqrt\Lambda)$
we arrive at:
\beq
\xi u^{''}(\xi)+2 u^{'}(\xi)=-~\xi~(u(\xi))^n[1+ \frac{2b}{n+1}
                            \xi ^2 u(\xi)] ~~,~~b= \frac{a}{\Lambda}
\label{difeq.2}
\eeq
The last equation is a non linear differential equation, which must be solved
for $\xi>0$, with the conditions $u(0)=1$ and $u^{'}(0)=0$. We also demand
that the solution remains positive, i.e. the solution drops from unity to zero.
 For $\alpha_s=0$ the density is a rapidly decreasing function of $\xi$.

 The above equation can be solved
analytically in the special case $a=0$ and only when $\lambda=-1/2$ (n=1)
or $\lambda=7/2$ (n=5). In these cases the solutions are:
\beq
\lambda=-1/2 \rightarrow~u(\xi)=\frac{sin\xi}{\xi}
\label{sol.1}
\eeq
and
\beq
\lambda=7/2 \rightarrow~u(\xi)=\frac{1}{(1+{\xi}^2/3)^{1/2}}
\label{sol.2}
\eeq
The  case $\lambda=-1/2$ may not have physical meaning, but it will serve as an
 illustrative example for the realistic cases to be discussed below.

 We should stress that the last solution associated with $\lambda=7/2$,
 admissible over all space, 
 can be considered as providing an adequate description of 
ordinary matter\footnote{The designation of $\lambda=7/2$ as representing
 ordinary matter is due to the work of Eddington ~\citep{EDDIN}, where he
 points
out that this value leads to the Plummer distribution. Furthermore this
distribution provides an adequate description of the galaxy (of ordinary
 matter) he was concerned with. A discussion of some of the thermodynamical
 aspects of the $\lambda=7/2$ distribution is also given in this paper.}.
In the absence of asymmetry no constraint is imposed on the parameter $\Lambda$
by the positivity and finiteness condition of the solution. The solution
given by Eq. (\ref{sol.1}) is constrained in the range $0\le \xi \le \pi$.
If we demand that the range of the potential for dark matter extends far
out in the halo, e.g. up to $x_r=r/r_s=20$,  we must have
$\Lambda=[\pi/{x_r}]^2=0.025$. This gives a useful constraint between
 $\Phi_0$ and $\rho_{\lambda}(0)$.

 The differential equation (\ref{difeq.2}) can, in general, be solved only
numerically. 
For $\lambda=1/2$ we find that there is a singularity at $\xi=12.43$, while the
first zero of the potential occurs at $\xi=4.35$. This leads to the constraint
$\Lambda=(4.35/20)^2=0.047$. For $\lambda=1$ the corresponding value is
$\Lambda=0.065$.

  The above results are significantly modified when the 
asymmetry parameter is turned on. One now can see that, depending on the
 parameters $\Lambda$ and $\alpha_s$, the shape of the density is significantly
modified.

We will examine the special case of $\lambda=7/2$, which corresponds to
 the Eddington solution for  ordinary matter. As we have seen in the absence
 of asymmetry the solution
can be found exactly. In the presence of asymmetry we distinguish two cases:

 1) Positive values of $\alpha_s$.

For small values of $b$ the solution has no roots. Beyond a
 critical value of $ b $ the solution attains the value zero. Then  further out
for still larger $\xi$ it
becomes negative. At some point it becomes singular. The extracted values of
$\Lambda$ range between $1.5$ and 0.12.

 2) Negative values of $\alpha_s$.

 In this
case  when the absolute value of $b$ becomes sufficiently large the solution
goes through zero. Again a value of $\Lambda$ can be extracted. The 
situation is quite unstable. Generally speaking for arbitrary values of $b$
the solution blows up 
 to infinity without  going through zero.
 One cannot extract values of $\Lambda$ in this case.

 In any case, these observations are intended as illustrations of
what is expected for dark matter where the exact solution cannot be obtained.
In the case of ordinary matter we will not consider $\alpha_s$
 different from
zero, since the range of the solution is small ( the density is assumed to
 vanish for $r\ge r_s$).

 Let us return to Eq. (\ref{distr.9}) and express it in terms of the velocity
 in our vicinity. We obtain:
\beq
 f(\upsilon)=~N[-2\Phi(r_s)-
\upsilon^2]^{\lambda} (1+\alpha_s\frac{\upsilon_t ^2}{\upsilon^2_m})
\label{dis.2}
\eeq
where $N$ is a normalization constant, which depends on $\lambda,\alpha_s$ and
$\upsilon_m$. From this we see that the maximum velocity in our vicinity ,
$\upsilon_m$, depends on the value of the potential, i.e.
\beq
\upsilon_m^2=-2~\Phi(r_s)=-2~\Phi_0~u(\sqrt{\Lambda}) 
\label{def.2}
\eeq
This means that
\beq
b=\frac{a}{\Lambda}=\frac{\alpha_s}{2~\Lambda~u(\sqrt{\Lambda})} 
\label{def.3}
\eeq
Our strategy is clear:

 1) Vary $b$ so that the solution has its first
zero at some value $\xi_0$ and is monotonically decreasing up to that point.

2)  Determine  $\Lambda$
 so that $\xi_0$ corresponds to a range $x_r$ of the potential $\chi(x)$.

3) From Eq. (\ref{def.3}) obtain an acceptable (self-consistent) value of 
$\alpha_s$, $\alpha_s=2 \Lambda b ~u(\sqrt {\Lambda})$.

4)  The parameter $\upsilon_m$ is obtained via the relation:
\beq
\upsilon_m^2=~=-2~\Phi_0~u(\sqrt{\lambda})=
            8\pi G_N r_s^2 \frac{u(\sqrt{\Lambda})~\rho_{\lambda}(0)}{\Lambda}  
\label{def.5}
\eeq
or equivalently
\beq
\upsilon_m=\upsilon_0~\sqrt{\eta_{\chi}} h_{\upsilon}(n,a,\Lambda)~~,~~
h_{\upsilon}(n,a,\Lambda)= [\frac{u(\sqrt{\Lambda})}
             {\Lambda ((u(\sqrt{\Lambda}))^n
             [1+ \frac{2~a}{n+1}~u(\sqrt{\Lambda})]}]^{1/2}~~,~~
                  n=\lambda+3/2
\label{def.6}
\eeq
where $\upsilon_0$ is our velocity of rotation around the center of our
galaxy (220 $km/s$).
 In obtaining the last equation we used our solution to relate the density
 $\rho_{\lambda}(0)$ to
 the density of dark matter in our vicinity, $\rho_0$, via the equation:
\beq
\rho_{\lambda}(0)=\frac{\rho_0}{((u(\sqrt{\Lambda}))^n[1+
                  \frac{2~a}{n+1}~u(\sqrt{\Lambda})]}~~,~~
                  n=\lambda+3/2
\label{def.7}
\eeq
The parameter $\eta_{\chi}$ is defined by $\eta_{\chi}=\rho_0/(0.3~GeV/cm^3)$.
 $\eta_{\chi}$ is normally taken to be unity \citep{Jungm}.

\begin{figure}
\includegraphics[height=0.8\textheight]{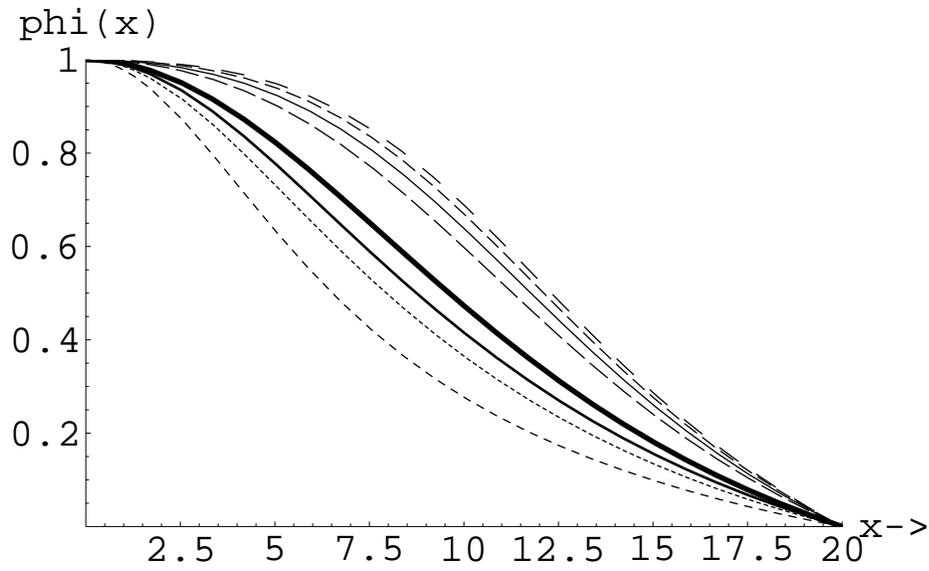}
\caption{The potential function in units of $\Phi_0$ for $\lambda=1/2$.
The notation of the curves is: Thick solid line,intermediate thickness
solid line, dotted line, fine dashing, long-fine dashing, 
intermediate-fine dashing and long-long dashing in the order of the
asymmetry of Table \ref{tab.1} (here $\alpha_s=0.000,-0.021,-0.038,-0.080,
0.062, 0.100,0.150,0.200$ and similarly for the other values of $\lambda$).
\label{f.phia}
}
\end{figure}

\begin{figure}
\includegraphics[height=0.8\textheight]{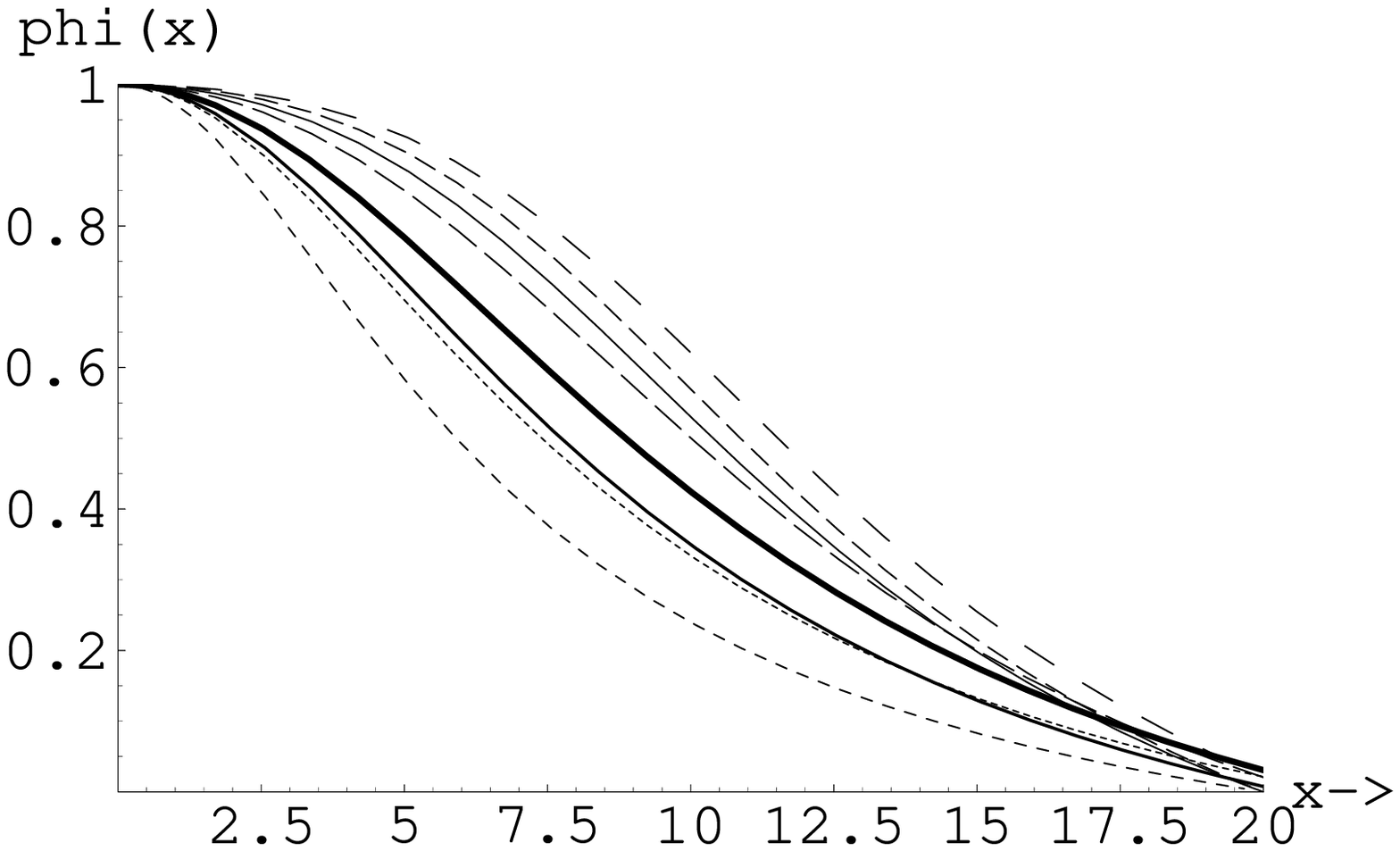}
\caption{The potential function in units of $\Phi_0$ for $\lambda=1$.
\label{f.phib}
}
\end{figure}

\begin{figure}
\includegraphics[height=0.8\textheight]{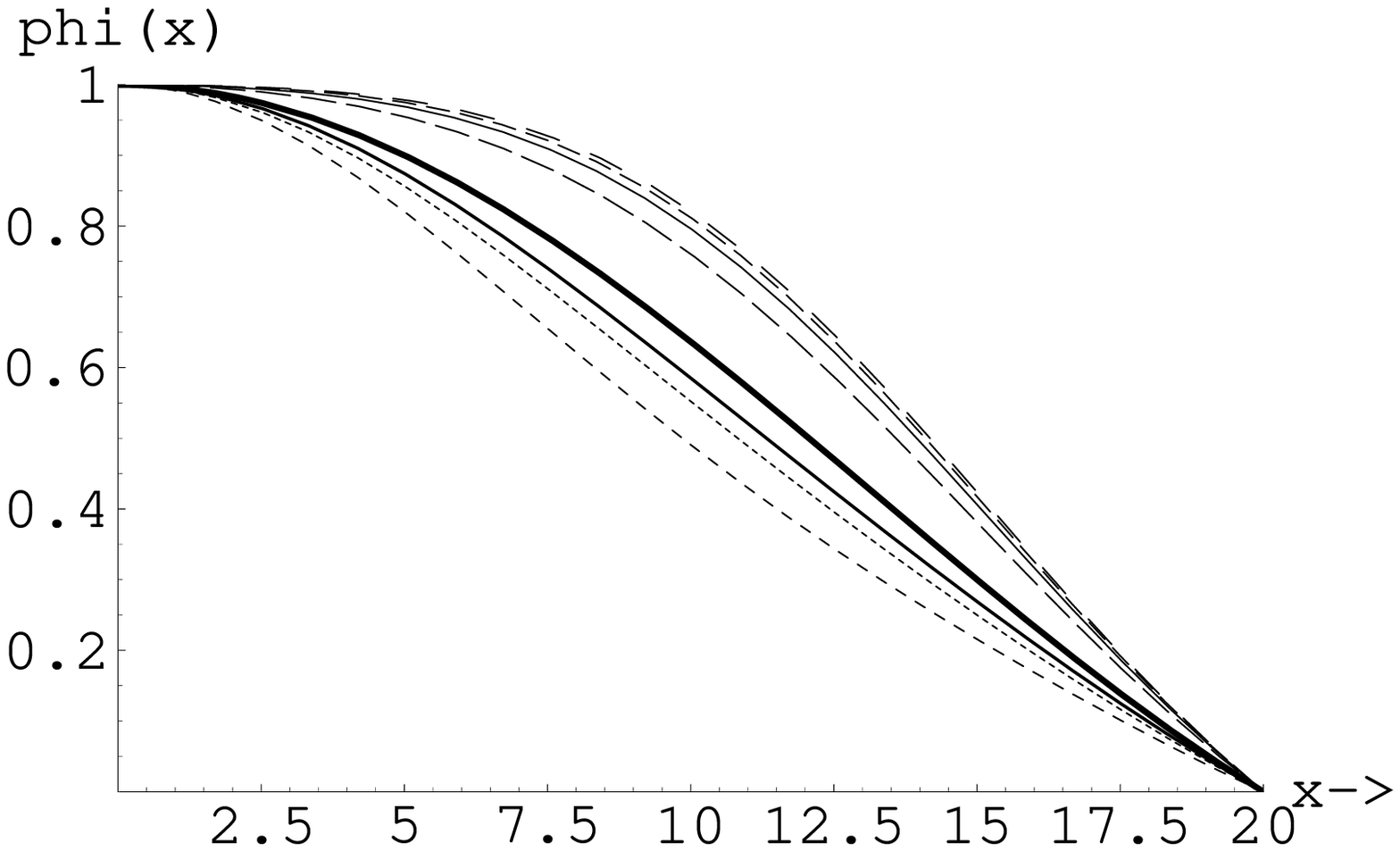}
\caption{The potential function in units of $\Phi_0$ for $\lambda=-1/2$.
\label{f.phic}
}
\end{figure}

\begin{figure}
\includegraphics[height=0.8\textheight]{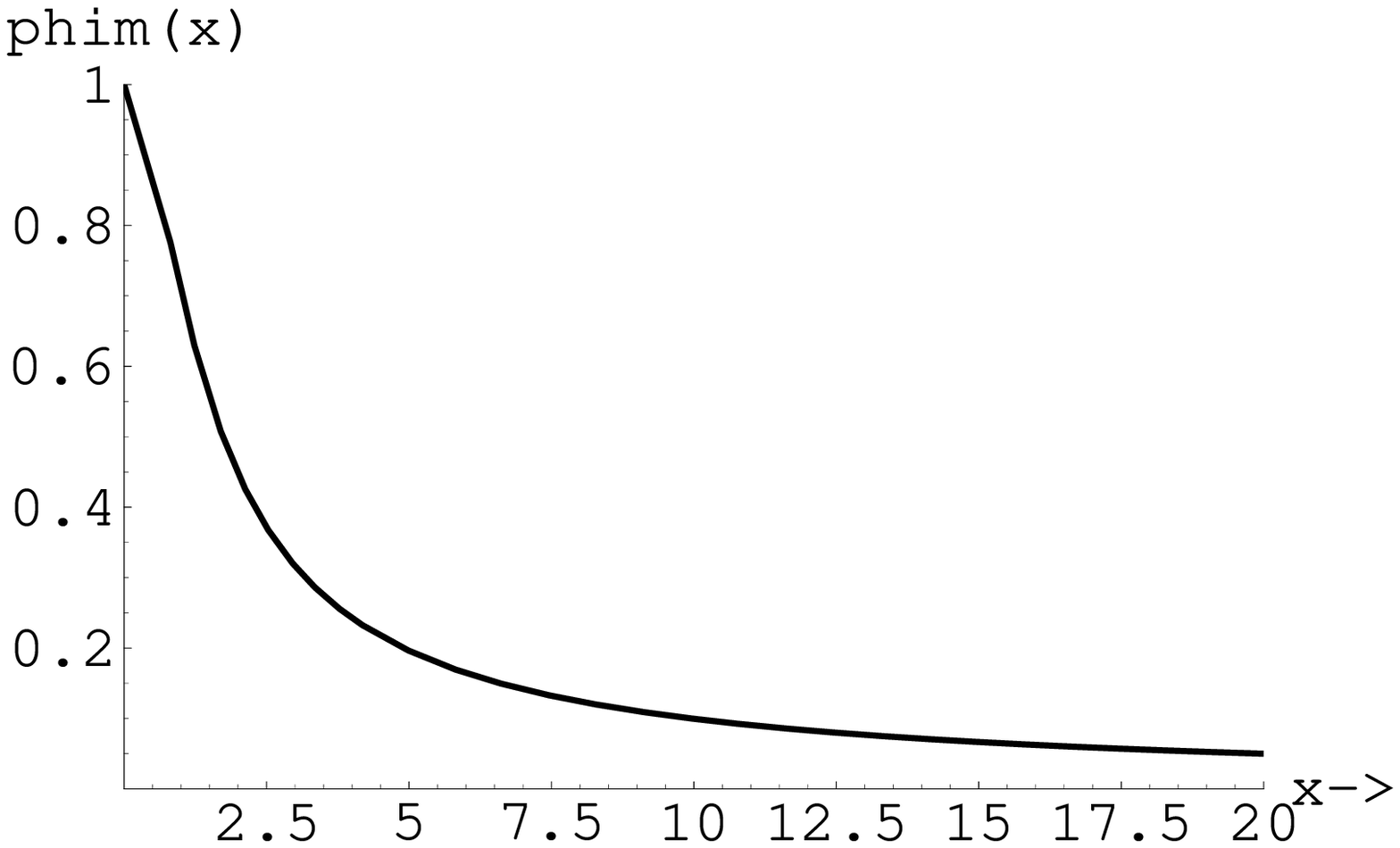}
\caption{The potential function in units of $\Phi_0$ for ordinary matter.
\label{f.phid}
}
\end{figure}

It is clear that in our treatment for a given $\lambda$ we have three
independent parameters, $b$, $x_r$ and $\rho_0$. The obtained results
for some interesting typical cases of $\alpha_s$ are shown in Table
\ref{tab.1}. The value of the parameter $h_{\upsilon}$ has the meaning
of $y_{esc}$ (see section 6) and it is rewarding that it is not far from
the value of 2.84 deduced phenomenologically \citep{Druk}. 
%%%%%%%%%%%%%%%%%%%%%%%%%%%%%%%%%%%%%%%%%%%%%%%%%%%%%%%%%%%%%%%%%%%%%
\begin{table}[t]  
\caption{
The various parameters describing the dark matter distribution for $x_r=20$
(for the definitions see text).
}
\label{tab.1}
%\begin{center}
\begin{tabular}{lrrrrr}
 & & & & &      \\
&  b & $\Lambda$ &a &$h_{\upsilon}$    &$\alpha_s$       \\
\hline
               & 0.00 &0.047 & 0.000 & 4.61  & 0.000      \\
               &-0.16 &0.064 &-0.010 & 3.97  &-0.021   \\
               &-0.24 &0.084 &-0.020 & 3.50  &-0.038   \\
               &-0.30 &0.134 &-0.040 & 2.79  &-0.080   \\
$\lambda=1/2$  & 1.45 &0.021 & 0.031 & 6.78  & 0.062   \\
               & 3.34 &0.015 & 0.050 & 8.02  & 0.100   \\
               & 6.95 &0.011 & 0.075 & 9.40  & 0.150   \\
               & 12.0 &0.008 & 0.100 &10.61  & 0.200   \\
\hline
               & 0.00 &0.062 & 0.000 & 4.05  & 0.000      \\
               &-0.13 &0.088 &-0.011 & 3.41  &-0.022    \\
               &-0.20 &0.102 &-0.020 & 3.18  &-0.040   \\
               &-0.27 &0.171 &-0.046 & 2.50  &-0.089   \\
$\lambda=1  $  & 0.70 &0.037 & 0.026 & 5.21  & 0.051   \\
               & 2.00 &0.026 & 0.052 & 6.13  & 0.103   \\
               & 4.20 &0.018 & 0.078 & 7.30  & 0.155   \\
               & 7.50 &0.013 & 0.099 & 8.50  & 0.201    \\
\hline
               & 0.00 &0.025 & 0.000 & 6.37  & 0.000      \\
               &-0.15 &0.032 &-0.005 & 5.57  &-0.010    \\
               &-0.25 &0.038 &-0.008 & 5.15  &-0.015    \\
               &-0.40 &0.050 &-0.012 & 4.51  &-0.025    \\
$\lambda=-1/2$ & 2.70 &0.010 & 0.025 &10.26  & 0.050    \\
               & 9.20 &0.005 & 0.050 &13.22  & 0.100    \\
               & 19.4 &0.004 & 0.075 &15.53  & 0.150    \\
               & 34.0 &0.003 & 0.101 &17.53  & 0.200    \\
 \hline
\end{tabular}
\end{table}
%%%%%%%%%%%%%%%%%%%%%%%%%%%%%%%%%%%%%%%%%%%%%%%%%%%%%%%%%
The potentials  obtained for the various such parameters
are shown in Figs \ref{f.phia} - \ref {f.phid}.
 It is clear that the shapes of these potentials are similar.

\begin{figure}
\includegraphics[height=0.8\textheight]{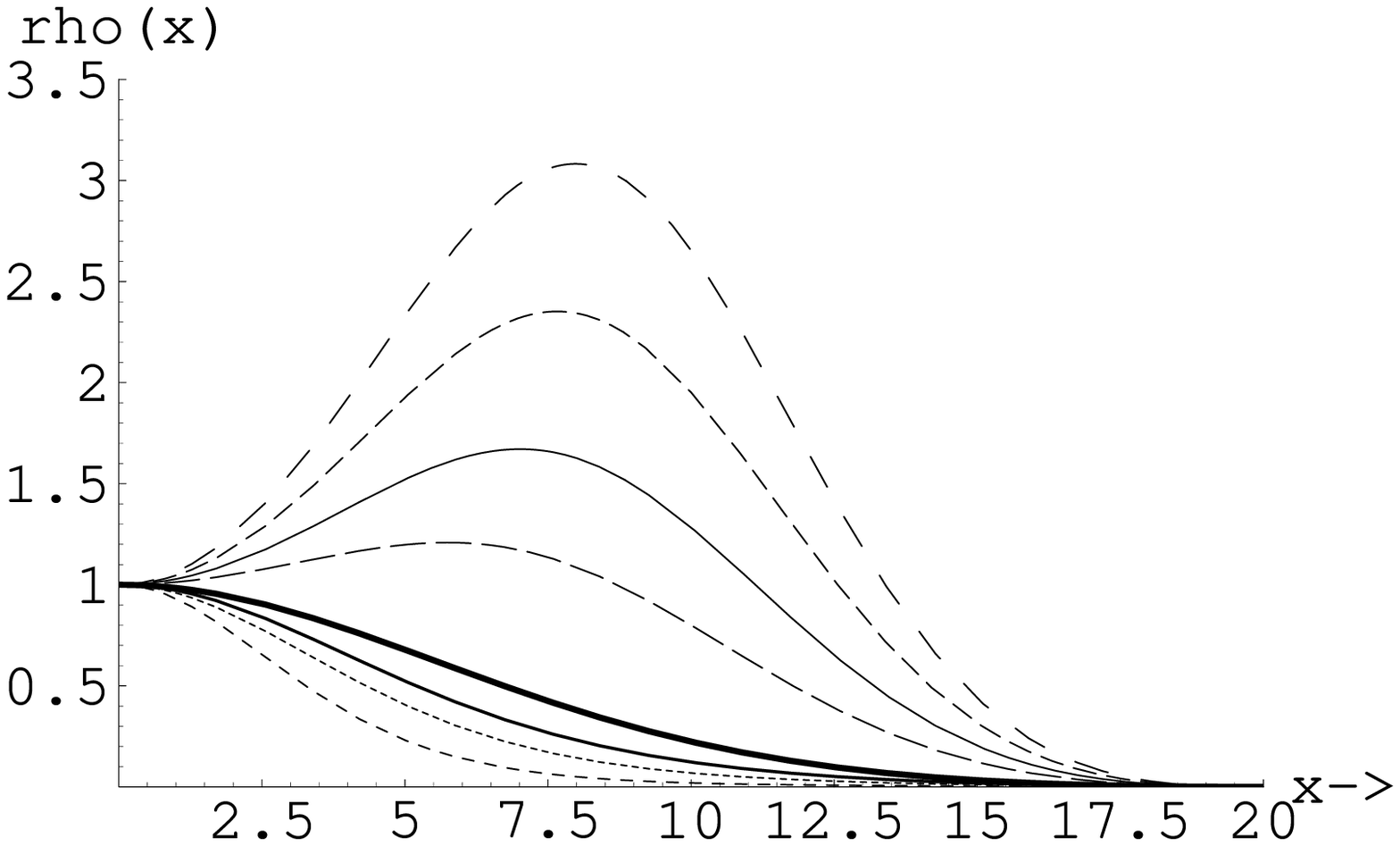}
\caption{The density in units of $\rho (0)$ for $\lambda=1/2$.
\label{f.rhoa}
}
\end{figure}

\begin{figure}
\includegraphics[height=0.8\textheight]{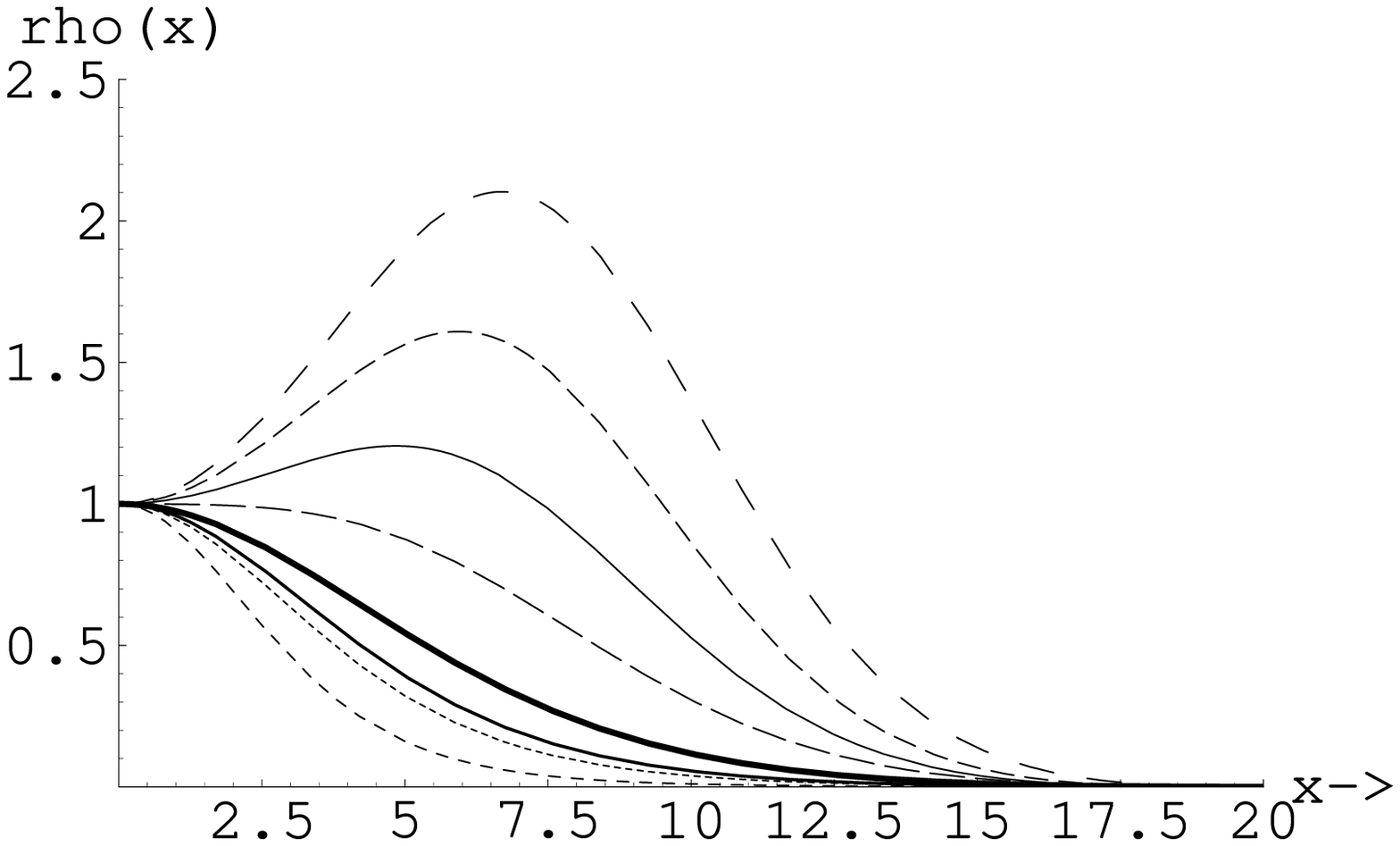}
\caption{The density in units of $\rho (0)$ for $\lambda=1$.
\label{f.rhob}
}
\end{figure}

\begin{figure}
\includegraphics[height=0.8\textheight]{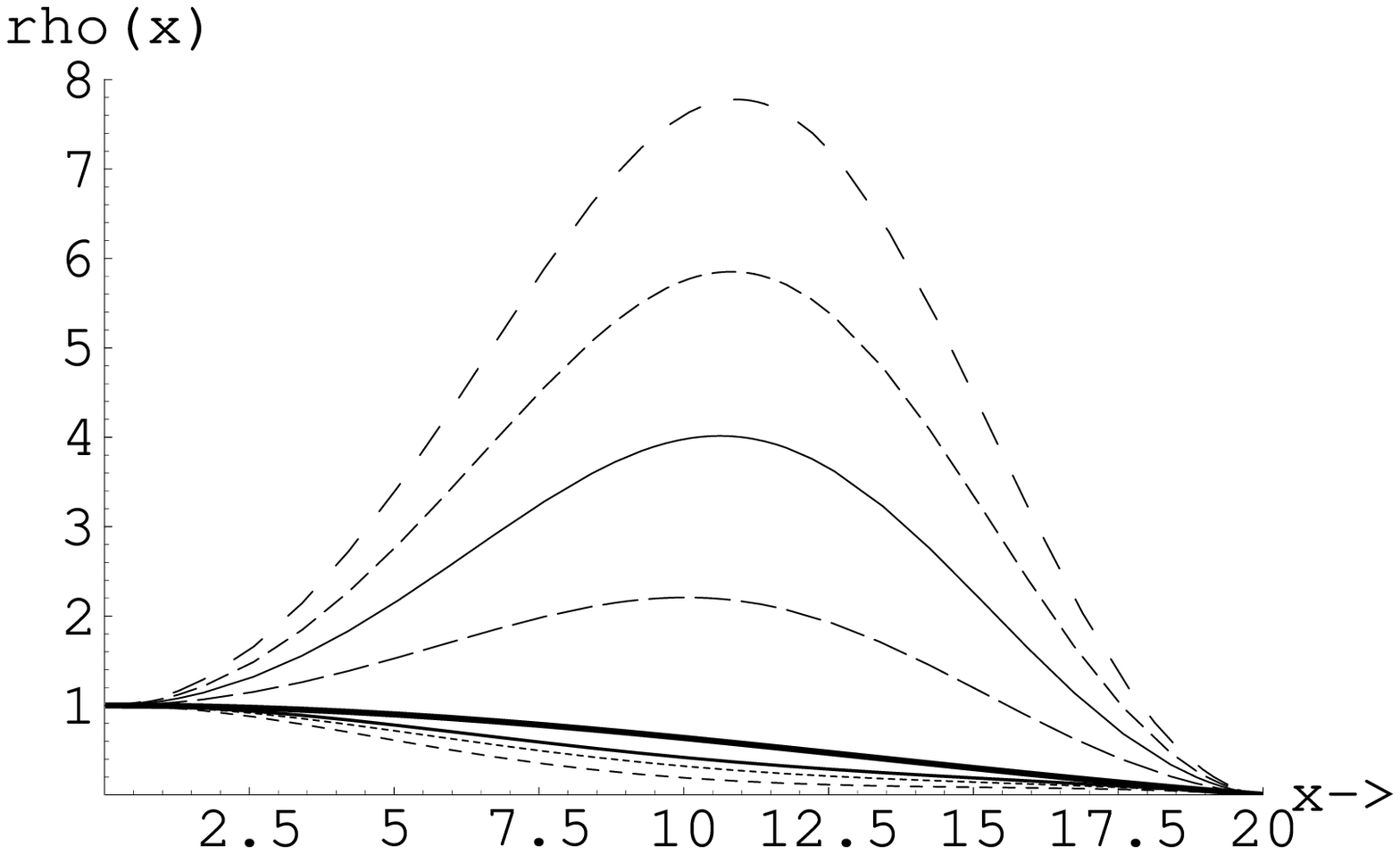}
\caption{The density in units of $\rho(0)$ for $\lambda=-1/2$.
\label{f.rhoc}
}
\end{figure}

\begin{figure}
\includegraphics[height=0.8\textheight]{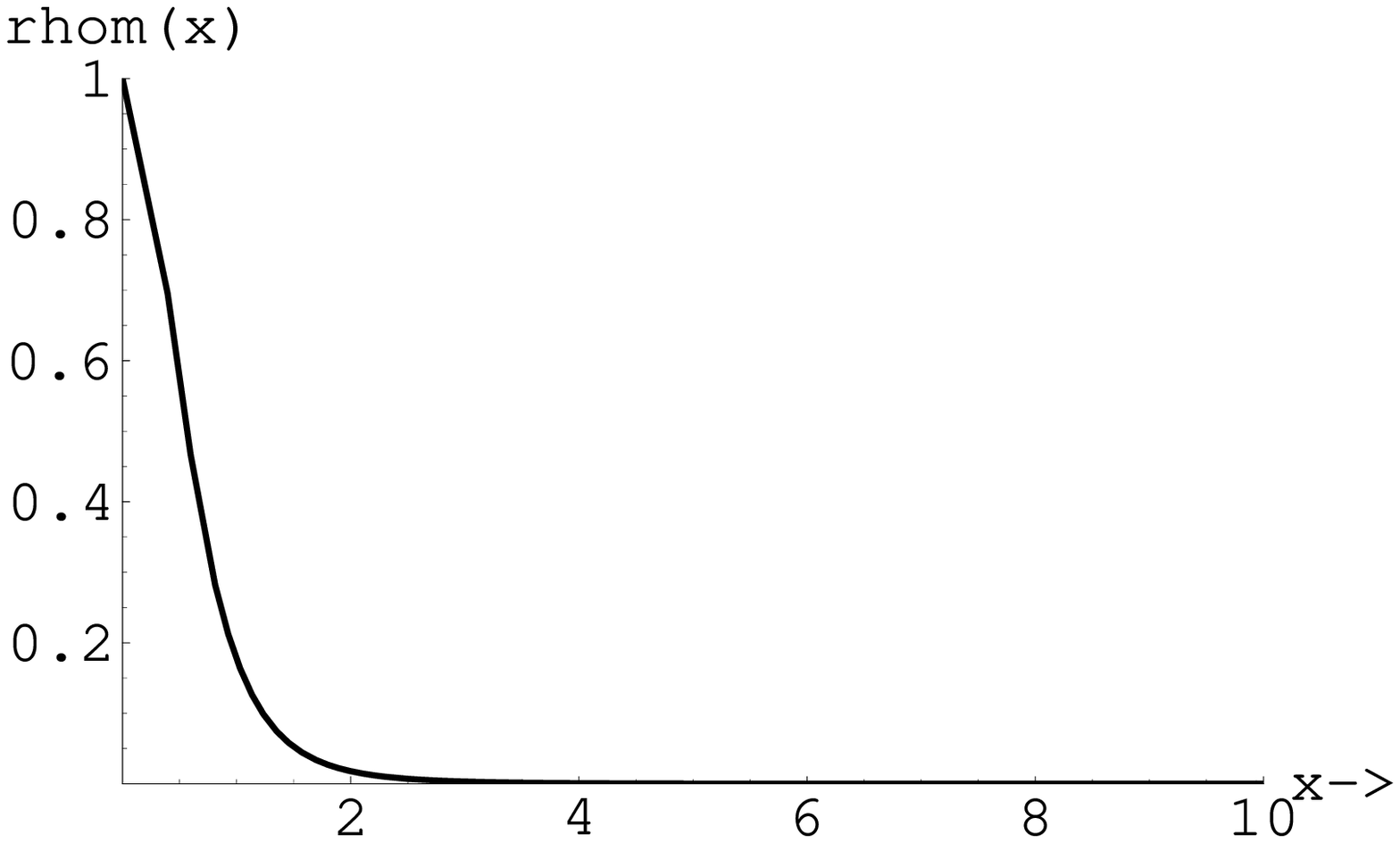}
\caption{The density in units of $\rho (0)$ for ordinary matter,
$\lambda=7/2$.
\label{f.rhod}
}
\end{figure}

The densities  obtained with the same  set of parameters
are shown in Figs \ref{f.rhoa} - \ref {f.rhod}.
 We observe that, even though the potentials are similar, the densities differ
substantially. The asymmetry parameter has a big effect on the  density,
especially for $\alpha_s$ greater than zero, when it changes substantially
around $x=8$. For $\alpha_s$ negative there is small  reduction of the density
 in the
region of interest, but, as we have already mentioned, only special values of
 $b$ are
 acceptable. A detailed rigorous mathematical study of the behavior of the 
solutions for $b$ negative is currently under study and it will appear 
elsewhere.

\section{Rotational Velocities}

\begin{figure}
\includegraphics[height=0.8\textheight]{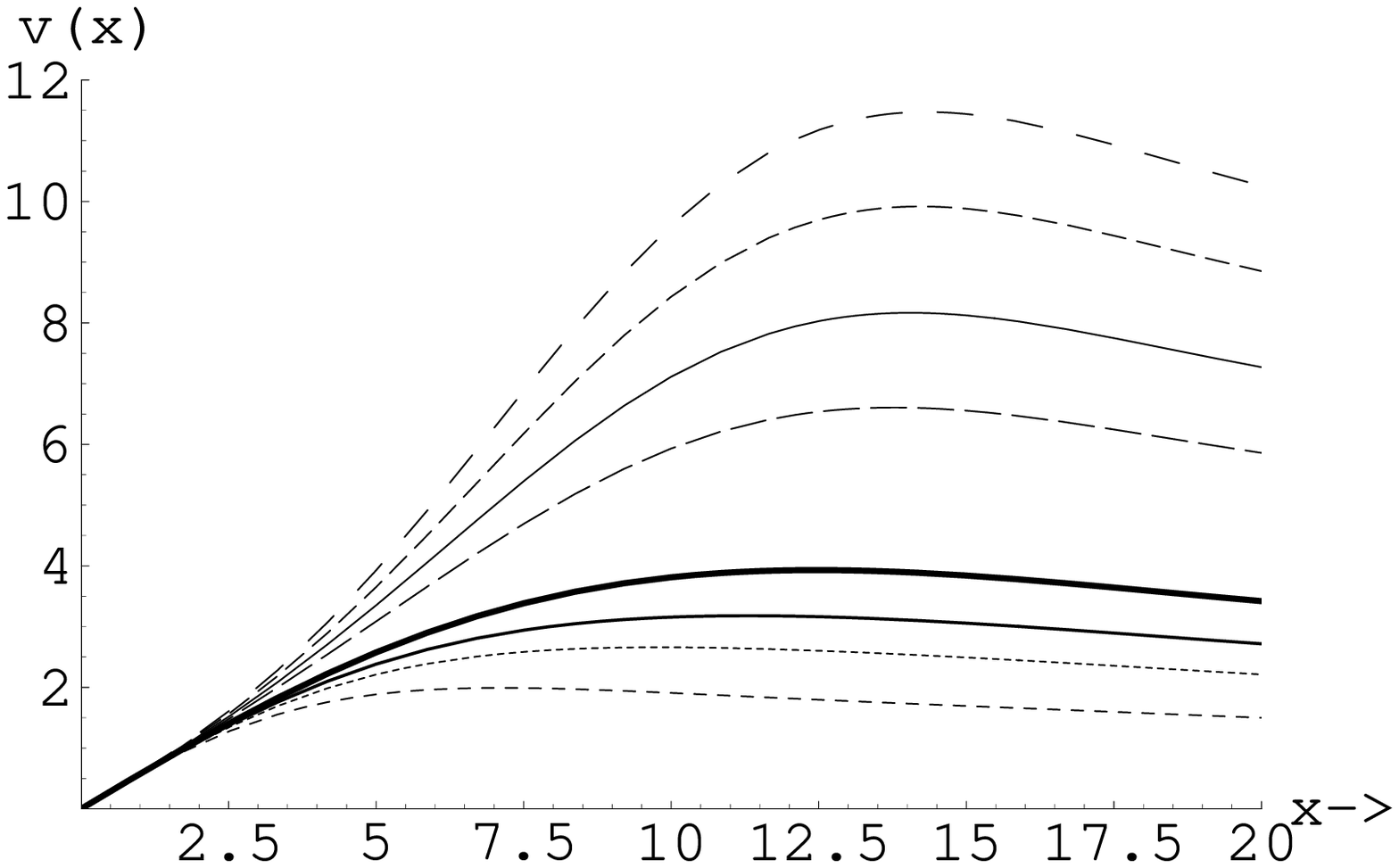}
\caption{The rotational velocities in units of $\frac{\upsilon_0}{\sqrt{2}}$
 for $\lambda=1/2$.
\label{f.va}
}
\end{figure}

\begin{figure}
\includegraphics[height=0.8\textheight]{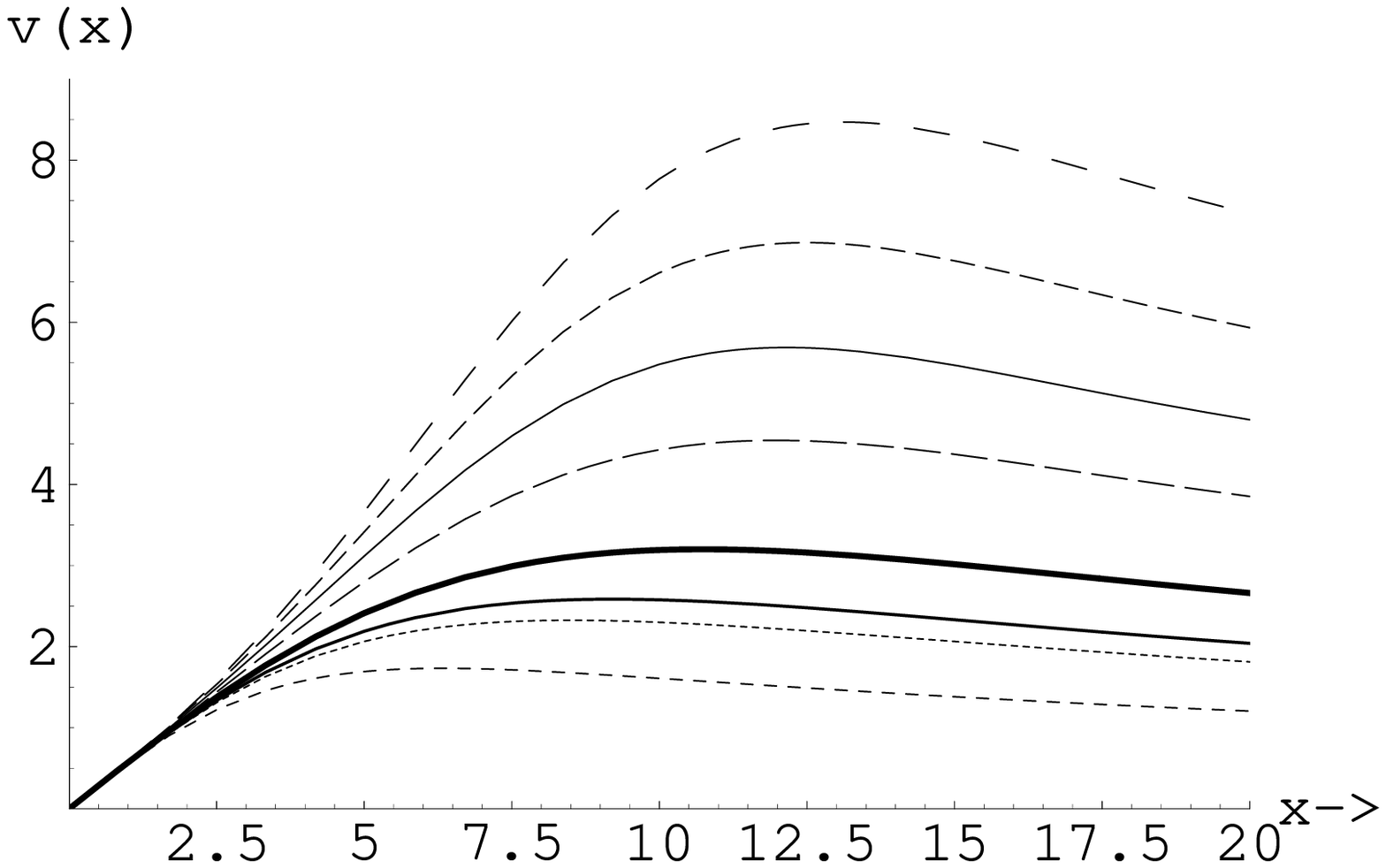}
\caption{The rotational velocities in units of $\frac{\upsilon_0}{\sqrt{2}}$
 for $\lambda=1$.
\label{f.vb}
}
\end{figure}

\begin{figure}
\includegraphics[height=0.8\textheight]{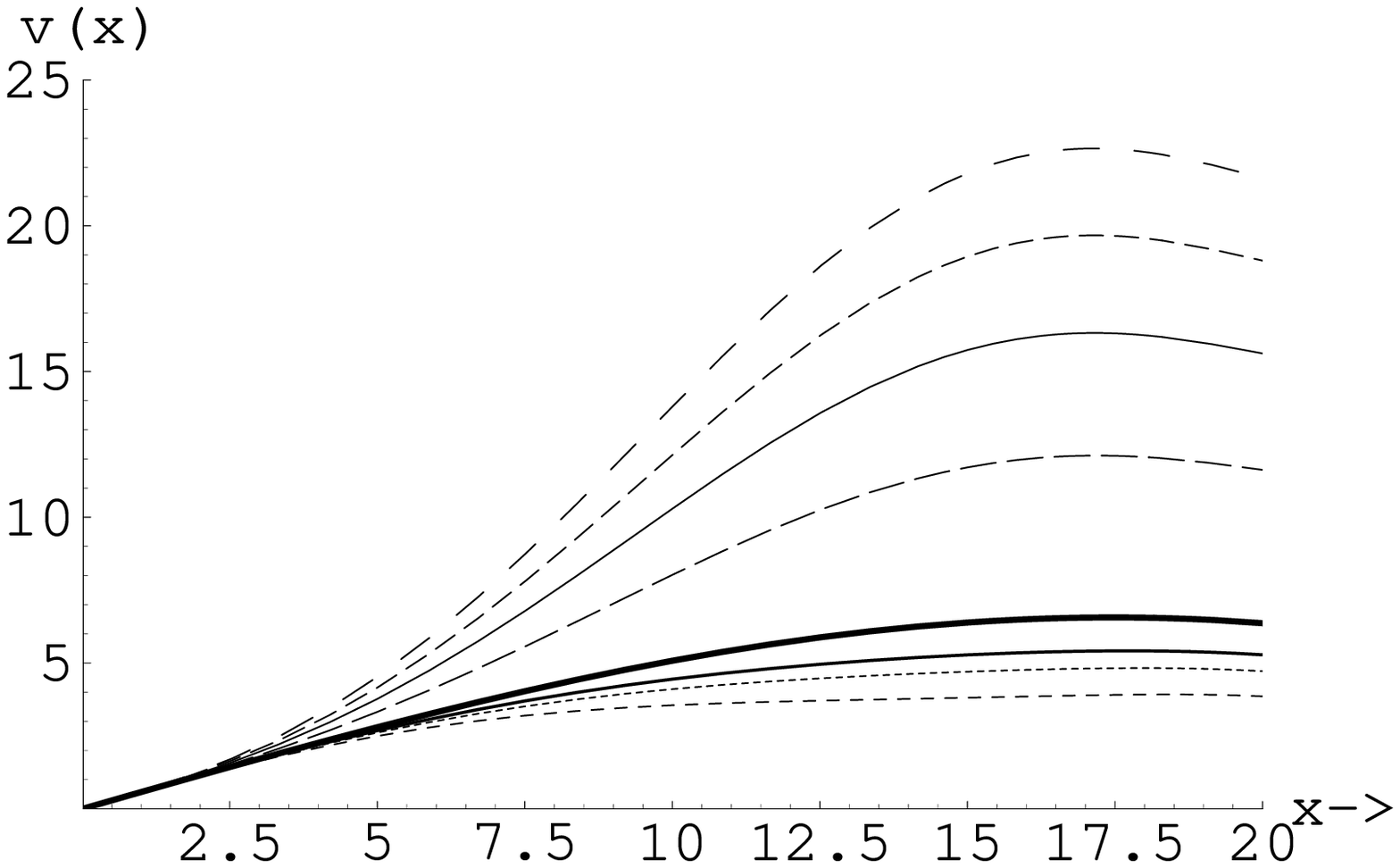}
\caption{The rotational velocities in units of $\frac{\upsilon_0}{\sqrt{2}}$
 for $\lambda=-1/2$.
\label{f.vc}
}
\end{figure}

 We are now in position to obtain the rotational velocity curves.
 The rotational velocity is given by
\beq
\upsilon_{\lambda}(x) = \frac{\upsilon_0}{\sqrt{2}} \sqrt{g_{\lambda}(x)}
\label{vel.1}
\eeq
with
\beq
g_{\lambda}(x)=\frac{1}{x} ~\int_0^x~ (x^{'})^2~dx^{'}~\psi_{\lambda}(x^{'})
\label{vel.2}
\eeq
 We see that the scale of the rotational velocity is set by
$ \frac{\upsilon_0}{\sqrt{2}}$ 
in  agreement with the data, see, e.g., the recent review \citep{Jungm}.
 For dark matter the above integrals can only be performed numerically. The
obtained rotational velocities are shown in Figs \ref{f.va} - \ref{f.vc}.
For ordinary matter ($\lambda=7/2)$ we make the choice of $\Lambda=3$ to get the
 familiar solution
\beq
\chi(x)=\frac{1}{(1+x^2)^{1/2}}~~,~~ \psi_{7/2}=\psi(x)=\frac{1}{(1+x^2)^{5/2}}
\label{vel.3}
\eeq
(We will drop the index $7/2$ and use the index $m$, for matter,
 when necessary).
Thus the rotational velocity due to ordinary matter is now given by
\beq
\upsilon_{m}(x) =\frac{ \upsilon_0}{\sqrt{2}}\sqrt{g_{m}(x)}
\label{vel.4}
\eeq
 For $x<1$ we write 
\beq
g_m(x)=g(x)
\label{vel.5}
\eeq
while for $x>1$  we have: 
\beq
 g_m(x) = \frac{1}{x}~g(1)
\label{vel.6}
\eeq
We now distinguish two cases:

i) Spherical galaxy.

 Using the above density we find:
\beq
 g(x) = \frac{1}{3}\frac{x^2}{(1+x^2)^{3/2}}
 \label{vel.5b}
\eeq

ii)  Spiral galaxy (disk).

\begin{figure}
\includegraphics[height=0.8\textheight]{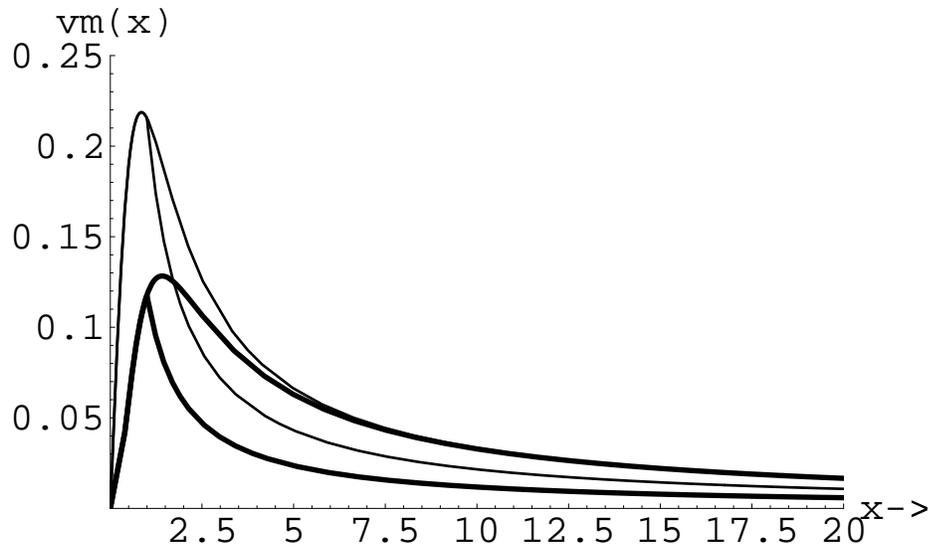}
\caption{The rotational velocities in units of $\frac{\upsilon_0}{\sqrt{2}}$
ordinary matter. Thick lines are associated with a spherical galaxy, while the
thin lines are associated with a spiral galaxy. The finite extent of the matter
density gives essentially the same results with that of infinite range. This
happens, because the density falls relatively fast.
\label{f.vd}
}
\end{figure}

In this case we will use the same density as above. In the Eddington theory
the equation, which relates the potential and the density is no longer of the 
above simple form. One  finds:
\beq
\frac{\Phi}{\Phi_0}=\frac{\Lambda}{12}
      [2~(\frac{\rho}{\rho (0)})^5-\ln[\frac{1-(\rho/ \rho (0))^5} 
      {1+(\rho/ \rho_0)^5}]] 
\label{eq.1}
\eeq
This complexity is of no concern to us since we do not need to use this
 complicated equation. In the case of ordinary matter we do not need to obtain
 the velocity distribution.
 Thus with the assumed density we find:
\beq
 g(x) = \frac{1}{3x}[1-\frac{1}{(1+x^2)^{3/2}}]
\label{vel.5a}
\eeq
 The above functions are plotted in Fig. \ref{f.vd}.
 In the graphs we show not only 
the physically interesting case, in which  the density vanishes
outside the radius of the galaxy, but, for illustrative purposes, the case
of infinite extent of the same density function . We see that, since the
density falls very fast, it makes very little difference, which form we use.

 By comparing the above
graphs of dark matter with those of ordinary matter we see that in the Eddington theory ordinary matter can make a significant contribution to the rotational
velocities, only if the density at the origin is much bigger compared to that
of dark matter. As we have mentioned above for ordinary matter there is no
 constraint between $\Phi_0$ and $\rho_0$. So we simply rescale ordinary matter
by a factor $C_{mdm}$ and write:

\beq
g_{matter}(x)=  C_{mdm} g_m(x)
\label{velm}
\eeq
the scaling factor can be determined by a comparison  to the experimental
 rotational curves.

\begin{figure}
\includegraphics[height=0.8\textheight]{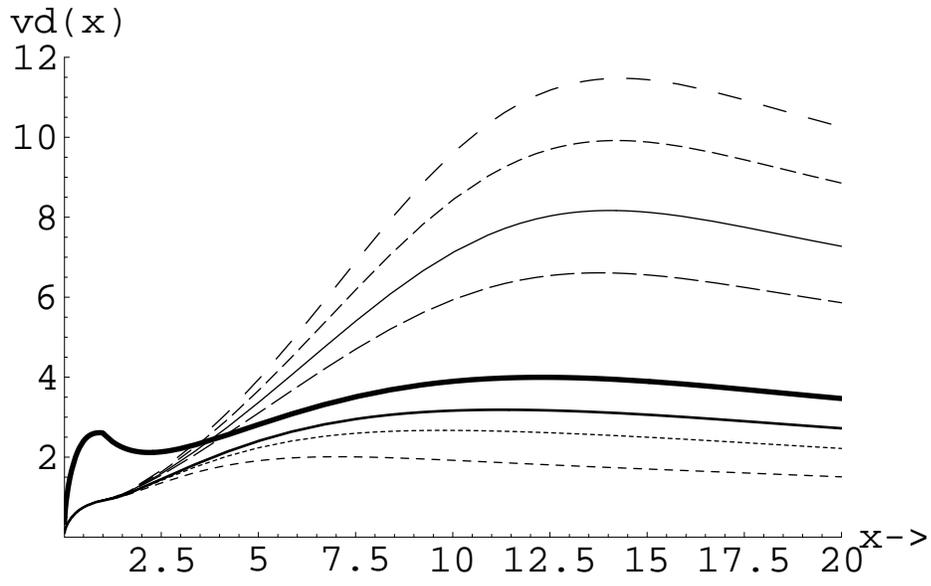}
\caption{The rotational velocities in units of $\frac{\upsilon_0}{\sqrt{2}}$
 for $\lambda=1/2$ in the case of spiral galaxy with $_{Cmdm}=30$.
\label{f.vd1}
}
\end{figure}

\begin{figure}
\includegraphics[height=0.8\textheight]{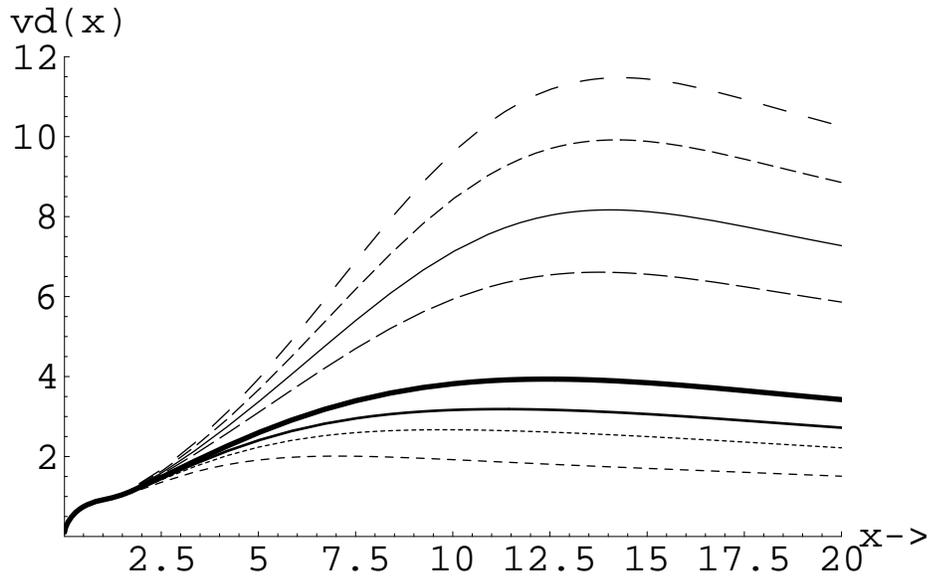}
\caption{The same as in Fig. \ref{f.vd1} for $C_{mdm}=2.5$.
\label{f.vd2}
}
\end{figure}

\begin{figure}
\includegraphics[height=0.8\textheight]{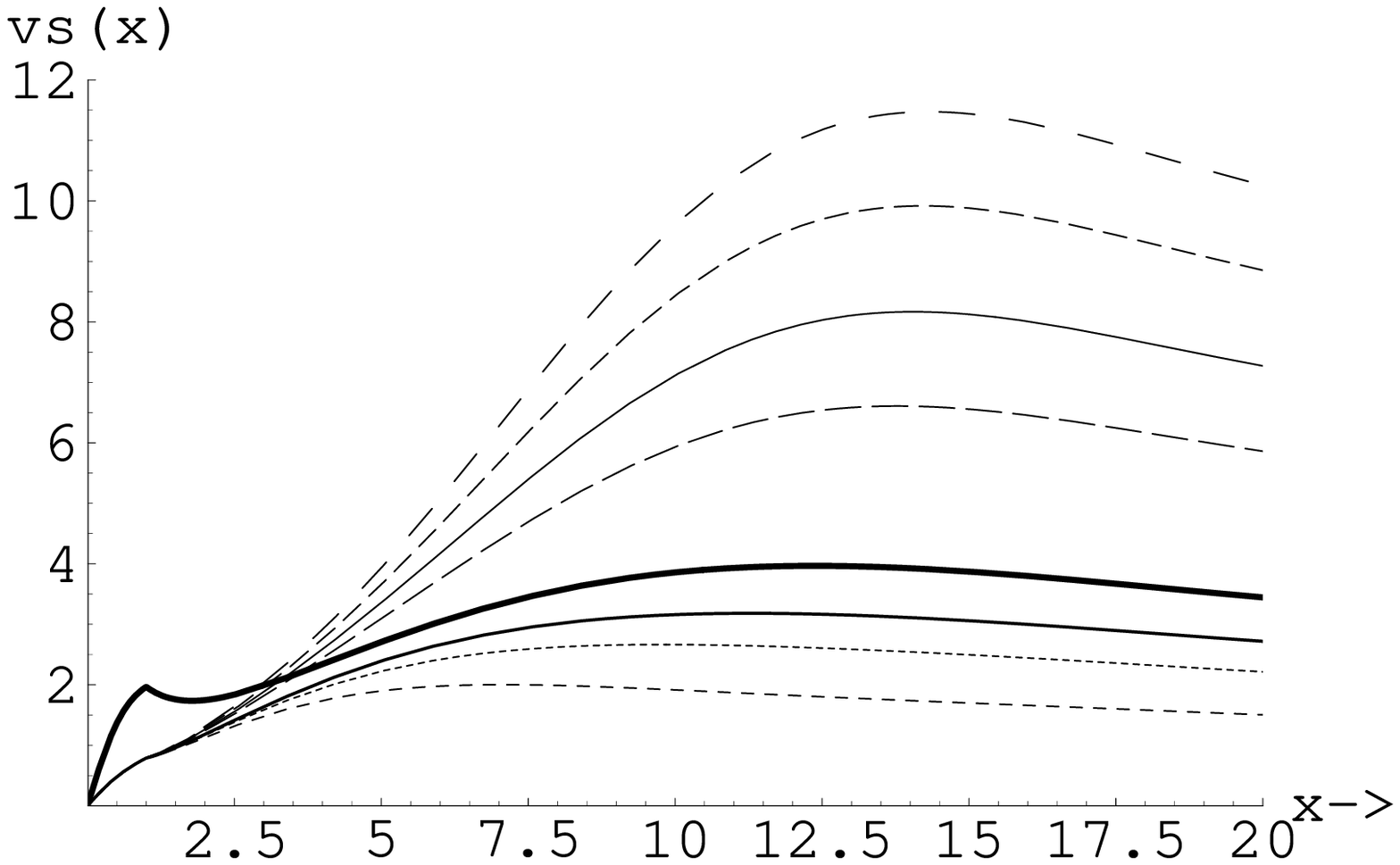}
\caption{The same as in Fig. \ref{f.vd1} for a spherical galaxy.
\label{f.vs1}
}
\end{figure}

\begin{figure}
\includegraphics[height=0.8\textheight]{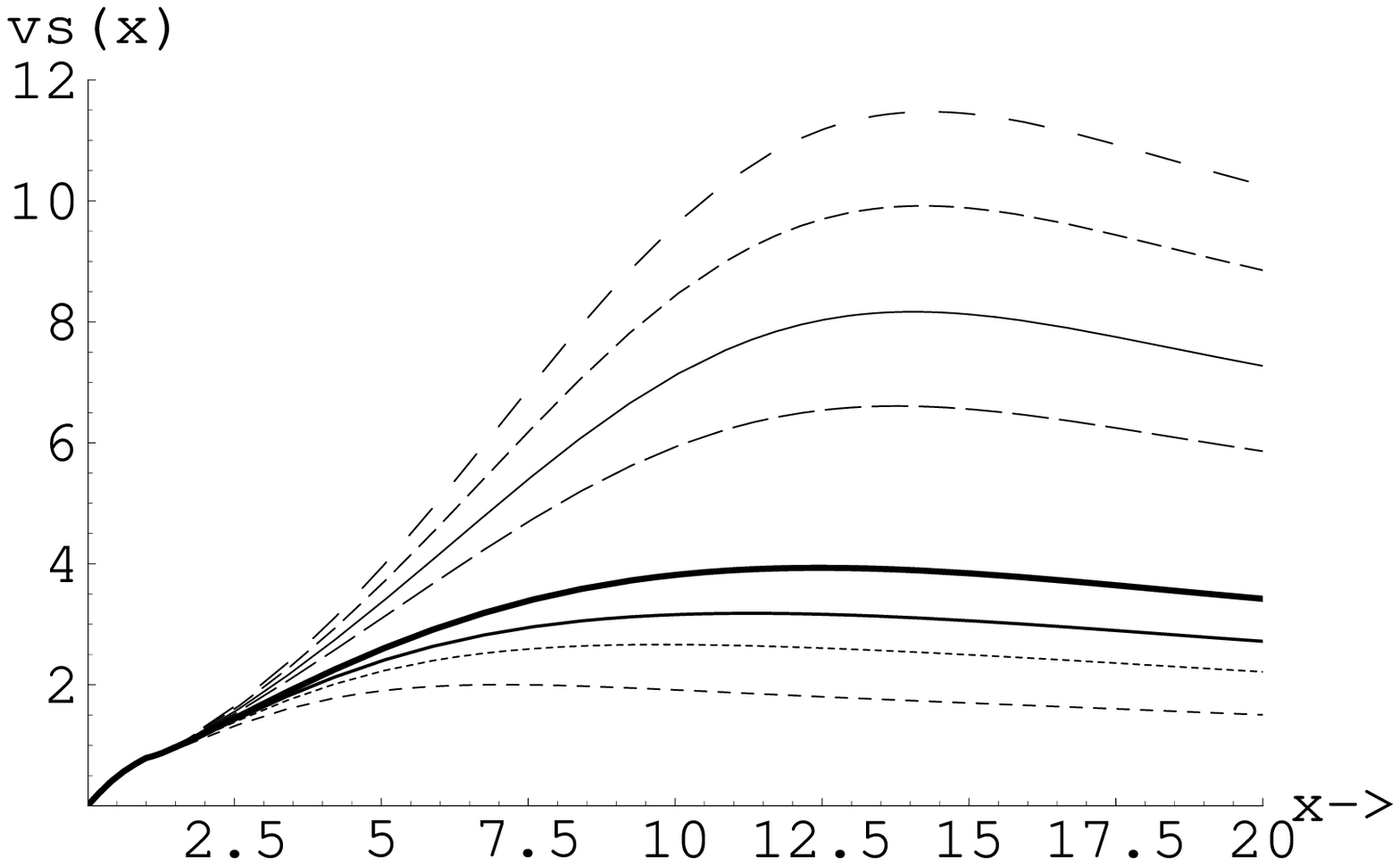}
\caption{The same as in Fig. \ref{f.vd2} for a spherical galaxy.
\label{f.vs2}
}
\end{figure}

Thus the rotational velocity due to both matter and dark matter is given by
\beq
\upsilon_{\lambda}(x)=\frac{\upsilon_0}{\sqrt{2}}
 \sqrt{g_{\lambda}(x)+g_{matter}(x)}
\label{vel.7}
\eeq
  The obtained rotational velocities for two values of $C_{mdm}$ are shown in
Figs \ref{f.vd1} - \ref{f.vs2}
 The larger value of $C_{mdm}$
seems to be in better agreement with the data, see, e.g., the recent review
\citep{Jungm}.

 It is clear from these curves that the presence of asymmetry has
a dramatic effect on the rotational velocities. Thus, in the context
of the Eddington theory as discussed here, a
 large asymmetry is excluded by the data on rotational velocities. A detailed
fitting of our input parameters to the rotational curves will not be attempted
here.

\section{Velocity distribution with respect to the galactic center}

 The above density via the Eddington formula leads to a velocity distribution
of the form:
\beq
 F(\upsilon,r)\propto~[-2\Phi(r)-
\upsilon^2]^{\lambda} (1+\alpha_s\frac{\upsilon_t ^2}{\upsilon^2_m})
\label{distr.11}
\eeq
The above velocities and the distance $r$ are defined with respect to the
center of the matter distribution, i.e to the center of our galaxy.
We note in the context of the Eddington theory the velocity distribution cannot
be Maxwellian. For a given distance it goes to zero at the boundaries of the
 corresponding ellipsoid.
It is customary to consider the value of the above distribution in our
vicinity, $r=r_s$. This way it reduces to the product of the local density and
the velocity distribution. The latter is given by Eq. (\ref{dis.2})
%\beq
% f(\upsilon)=~N[-2\Phi(r_s)-
%\upsilon^2]^{\lambda} (1+\alpha_s\frac{\upsilon_t ^2}{\upsilon^2_0})
%\label{dis.2}
%\eeq
where $N$ is a normalization constant, which depends on $\lambda,\alpha_s$ and
$\upsilon_m$. 
The above notation was introduced to make the last equation coincide with the
standard expression when the function f is chosen to be Maxwellian, i.e.

$Exp(-\frac{\upsilon^2-\alpha_s\upsilon^2_t}{\upsilon_0^2}) \rightarrow$
$ (1+\alpha_s\frac{\upsilon_t^2}{\upsilon^2_0})$
$Exp(-\frac{\upsilon^2}{\upsilon^2_0})$

 for sufficiently
small $\alpha_s$. In this limit we see that $\alpha_s$ coincides with the
parameter $-\lambda$ of Vergados \citep{Verg00} {\it et al} \cite{Druk}
(in the present work $\lambda$ is used for another purpose). 

 It is
straightforward to find that the normalization factor $\tilde{N}$ is given by
\beq
\tilde{N}^{-1}(\lambda,\alpha_s,\upsilon_m)=
2\pi \upsilon_m^{2\lambda+3}\beta(\lambda+1,3/2)
    [1+\frac{4}{3}~\alpha_s~ \frac{\beta(\lambda+1,5/2} {\beta(\lambda+1,3/2})]
\label{norm.1a}
\eeq
with $\upsilon_m$ given by Eq. \ref{def.2}. 
 In the special case of dark matter ($\lambda=1/2)$ it becomes
\beq
\tilde{N}^{-1}(1/2,\alpha_s,\upsilon_m)= \frac{\pi^2}{4} \upsilon_m ^4 
     [1+\frac{1}{3} \alpha_s ]
\label{norm.1b}
\eeq
 From the above formulas we see that the velocity of dark matter with respect
to the galactic center ranges from $0$ to a maximum speed
$\upsilon_m=|2\Phi(r_s)|^{1/2}$.  Since the escape velocity is determined by
the potential $\Phi(r_s)$ due to all kinds of matter, the velocity $\upsilon_m$
is not simply related to $\upsilon_{esc}$.
%The escape velocity is assumed to be given by $\upsilon_{esc}=2.84~\upsilon_0$.
 Thus, since the
distribution function must remain positive, if $\alpha_s<0$ its  absolute
 value is bounded.
 This imposes a constraint, since in
the traditional analysis with only axially symmetric Gaussian distribution
it leads to negative $\alpha_s$, i.e. enhanced dispersion in the galactocentric
 direction, a
 phenomenologically preferred result \citep{Druk}.
 The data on rotational curves may provide an additional constraint on  the
 negative values of $\alpha_s$.
 For positive values of $\alpha_s$ the constraint coming from
the rotational curves is, as we have seen, more stringent.

From then on one proceeds in the usual way to obtain the velocity distribution
with respect to the laboratory.

\section{Velocity distribution with respect to the laboratory}

For this transformation one needs the velocity of the sun around the galaxy
$\upsilon_0 = 220 Km /s$, a fraction of the escape velocity, which is
$\upsilon_{esc}=625Km/s=2.84~\upsilon_0$ \citep {Druk}.

It is convenient to choose coordinate system with its  polar z-axis in the the
 direction of the
disc's rotation, i.e. in the direction of the motion of the the sun,
 the x-axis is in the outward radial direction and  
 the y-axis is perpendicular to the plane of the galaxy.
Since the axis of the ecliptic \citep{KVprd}. 
lies very close to the $y,z$ plane the velocity of the earth around the
the center of the galaxy is given by 
\beq
 \mbox{\boldmath $\upsilon$}_E \, = \, {\bf \upsilon}_0 {\bf \hat z}\, 
                      + \, \mbox{\boldmath $\upsilon$}_1 \,
= \, {\bf \upsilon}_0 {\bf \hat z}~+ \upsilon_1(\, \sin{\alpha} \, {\bf \hat x}
-\cos {\alpha} \, \cos{\gamma} \, {\bf \hat y}
+\cos {\alpha} \, \sin{\gamma} \, {\bf \hat z} \,)
\label{3.6}  
\eeq
where $\alpha$ is the phase of the earth's orbital motion, $\alpha =2\pi 
(t-t_1)/T_E$, where $t$ is the time of observation and $t_1$ is around second
 of June and $T_E =1$ year. The
magnitude of the Earth's velocity is much smaller than that of the sun,
i.e. $\delta_1=2 \upsilon_1 / \upsilon_0 = 0.27$
 The velocity of the earth around its own axis is even smaller and it is
usually neglected.

One can now express the above distribution in the laboratory frame 
by writing $ 
 \mbox{\boldmath $\upsilon$}^{'}=
 \mbox{\boldmath $\upsilon$}
 \, + \, \mbox{\boldmath $\upsilon$} _E \,$ ,
where the prime indicates the velocity with respect to the center of the galaxy.
Indicating by $y$ the velocity of the LSP (Lightest Supersymmetric Particle)
in units of $\upsilon_0$, i.e. by
defining $y=\upsilon/\upsilon_0$, we find
\beq
f(y,\theta,\phi)=N(\lambda,\alpha_s,y_m)~[y_m^2~-~Y(y,\theta,\phi)]^{\lambda}
             [1+\alpha_s(Y(y,\theta)-
                     (y~sin\theta~\cos\phi-\frac{\delta_1}{2}~\sin\alpha)^2)]
\label{distr.12}
\eeq
with
\beq
N(\lambda,\alpha_s,y_m)= \upsilon_0^{2\lambda+3}
\tilde{N}(\lambda,\alpha_s,\upsilon_m)
\label{norm.1e}
\eeq
 where $\tilde{N}(\lambda,\alpha_s,\upsilon_m)$ is given by Eq. \ref{norm.1a},
 $y_m=\upsilon_m / \upsilon_0$ and
\barr
Y(y,\theta,\phi) &=&1+\frac{\delta_1^2}{4} +y^2
 +2y \cos \theta +\delta_1[y~\cos\theta~\cos\alpha~\sin\gamma
\nonumber \\ &-&
     y~\sin\theta~\sin\phi~\cos\alpha \cos\gamma
     +y~\sin\theta~\cos\phi~\sin\alpha]
\label{distr.13}
\earr
with 
%$\delta _1= \frac{2 \upsilon_1}{\upsilon_0}=0.27$ ,
% $y=\frac{\upsilon}{\upsilon_0}$ and $y_m=\frac{y_{esc}}{\sqrt{2}}$~
In the conventional axially symmetric Gaussian velocity distribution this function
 is given by
\barr
f(y,\theta,\phi)& = &\frac{N(\alpha_s,y_{esc})}{\pi\sqrt{\pi}}~
                 Exp[-(\alpha_s +1)Y(y,\theta,\phi)\\
\nonumber
          &+& \alpha_s(y~sin\theta~\cos\phi-\frac{\delta_1}{2}~\sin\alpha)^2)]
\label{distr.14}
\earr
In this case $0~\leq y \leq~y_{esc}$, but the upper cutoff is introduced here
 artificially. The normalization here is defined so that 
$N(\alpha_s=0,y_{esc}\rightarrow\infty)=1$ \citep{Verg00}

The detection rate in direct dark matter experiments is obtained by convoluting
the the relevant cross section
with the above velocity distribution. If the dark matter candidate is the LSP,
the $\alpha$-dependence of the above distribution
, present only when $\delta_1 \neq 0$,
gives rise to the modulation effect, i.e. the dependence of the rate on the
Earth's motion. This signal can be used to discriminate against background.
%Results obtained this way will appear elsewhere.

\subsection{The non directional rate}

The non directional differential event rate is given by: 
\beq
 \frac{dR}{du}=\bar{R} \sqrt{\frac{2}{3}}T(u)~,~T(u)=a^2|F(u)|^2 \Psi(a\sqrt{u})
\label{T.1}
\eeq
for the coherent mode and
\beq
 \frac{dR}{du}=\bar{R} \sqrt{\frac{2}{3}}T_{spin}(u)~,~
       T_{spin}=a^2|F_{11}u)| \Psi(a\sqrt{u})
\label{T.2}
\eeq
where $\bar{R}$ and $\bar{R}_{spin}$ are the total rates for the coherent
 and the spin contributions associated with an average LSP velocity,
 $\sqrt{<\upsilon^2>} =\sqrt{(3/2)}\upsilon_0$. These parameters, which
 carry the 
dependence on the SUSY parameters, are the most important ones, but they
 are not of interest in our present calculation.
$F(u)$ is the form factor, entering the coherent scattering
 and $F_{11}(u)$ is the spin response function entering via the axial current
\citep{JDV02}. The function $\Psi$ depends on the LSP distribution velocity
employed and is a function of the energy $Q$ transferred to the nucleus
\beq
u=\frac{Q}{Q_0}~~,~~Q_{0}=4.1\times 10^{4}~A^{-4/3}~KeV
\label{u.1}
\eeq
 where $A$ is the nuclear mass number and the parameter $a$ is given by:

\beq
a=[\sqrt{2}\mu_r b \upsilon_0]^{-1}
\label{a.1}
\eeq
where $\mu_r$ is the reduced mass of the LSP-nucleus system and $b$ is the
(harmonic oscillator) size parameter.

 The function, which is basic to us, $\Psi$, is given by 
\beq
\Psi(x)=~\int_0^x~ dy~\int_0^{\pi} ~ \sin\theta d\theta \int_0^{2\pi}~d\phi~y~
        f(y,\theta,\phi)
\label{psi.1}
\eeq
with $0\leq x \leq y_m-1+(\delta_1/2)\cos\alpha~\sin\beta$.

The total rate is given by:
\beq
R= ~\int_{u_{min}}^{u_{max}}\frac{dR}{du}du
\label{tot.1}
\eeq
where $u_{min}$ is determined from the cutoff energy of the detector and 
$u_{max}=(y_m/a)^2$

From now on we will specialize our results to the case $\lambda=1/2$, which
yields an adequate description of the rotation curves in the case of dark
 matter.  In the case of this velocity distribution,
 unlike the Gaussian one, one cannot 
approximate the distribution by a power series in $\delta_1$.
 The reason is that there may be threshold problems, when the argument of the
 square root approaches zero.
 To simplify matters we will still 
make use of the fact that the velocity of the Earth around the Sun is much
smaller than the velocity of the Sun around the galaxy, $\delta_1<<1$. So, if
we expand the previous expression into a Fourier series with respect to
the phase of the Earth, $\alpha$, only the lowest terms will become important.
 In other words to leading order in $\delta_1$ it can be  put in the
 form:
\beq
R= \bar{R}[R_0+(R_1~\cos\alpha~\sin\gamma~-~R_2~\cos\alpha~\cos\gamma~+
~R_3~\sin\alpha)\delta_1/2]
\label{tot.2}
\eeq
It turns out that the expansion coefficients $R_2$ and $R_3$ are zero. We can
thus conveniently fit the rate with the formula:
\beq
R= \bar{R}~t~[1+h~\cos\alpha]
\label{tot.3}
\eeq
where $h$ is the modulation amplitude (the difference between the maximum and
the minimum is equal to $2|h|$).

\begin{figure}
\includegraphics[height=0.3\textheight]{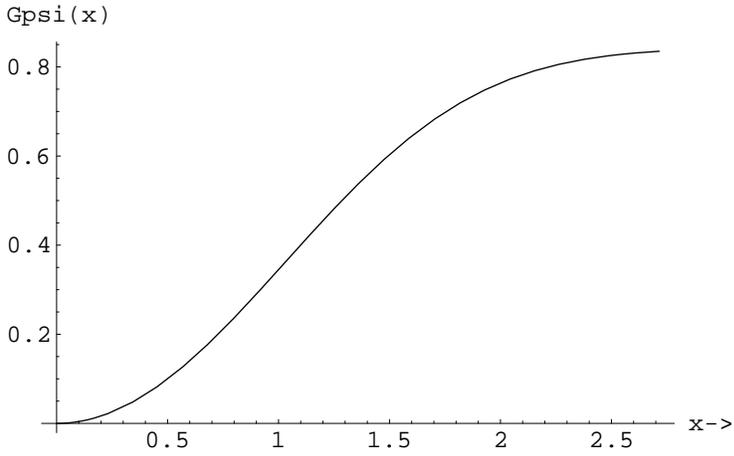}
\caption{The function $\Psi(x)$ for dark matter in the case 
of the symmetric Gaussian distribution.
\label{psi1}
}
\end{figure}

\begin{figure}
\includegraphics[height=0.3\textheight]{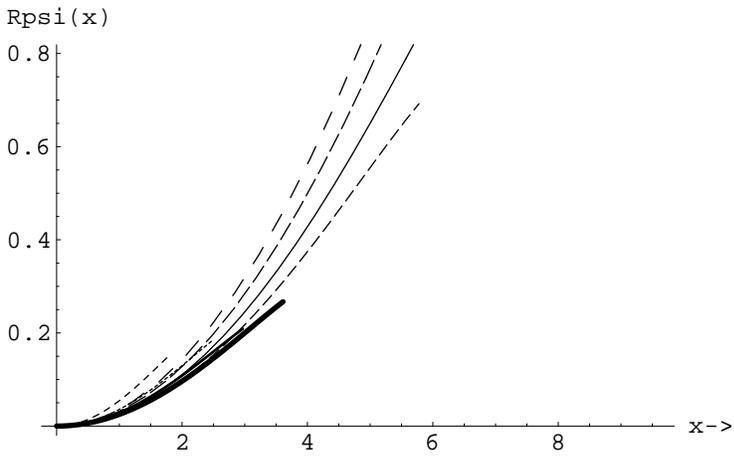}
\caption{The function $\Psi(x)$ in the case of dark matter for the choice
 $\lambda=1/2$
of the Eddington theory. The graphs have been labeled as in Fig. \ref{f.va}.
\label{psi2}
}
\end{figure}

In the case of no modulation, $\delta_1=0$, the angular integrals can be done
analytically to yield:
%\beq
%\Psi(x)=2 \pi N(\alpha_s,y_m)[J_{even}(x)+J_{odd}(x)]
%       \sum_{n=3,5,7}~A_n(\alpha_s,y_m)\frac{1}{n}J_n(x,y_m)
%\label{psi.2}
%\eeq
\beq
\Psi(x)=2 \pi N(1/2,\alpha_s,y_m)
        [\frac{(1+\alpha_s y_m^2)}{3}J_3(x,y_m)- \frac{\alpha_s}{20}J_5(x,ym)]
\label{psi.2a}
\eeq
%\beq
%J_{odd}(x)=[(y_m^2-(x-1)^2)^{7/2} +(y_m^2-(x+1)^2)^{7/2}
%-2(y_m^2-1)^{7/2}]\frac{\alpha_s}{14}
%\label{psi.2b}
%\eeq
with $0\le x \le y_m-1$ and
\beq
J_n(x,y_m)=J_{int}(n,y_m,x-1)-J_{int}(n,y_m,x+1)+2~J_{int}(n,y_m,1)
\label{psi.3}
\eeq
% with $A_n(\alpha_s,y_m)=(1-(\alpha_s/2)y_m^2,\alpha_s/4,\alpha_s/4)$ and
\beq
J_{int}(n,y_m,y)=\int_0^y [y_m^2-z^2]^{n/2}dz
\label{psi.4}
\eeq
The above integral can be done analytically to yield:
\beq
J_{int}(n,y_m,y)=y(y_m)^n~
                _2F_1(\frac{1}{2},\frac{-n}{2},\frac{3}{2},\frac{y^2}{y_m^2})
\label{psi.5}
\eeq
where $_2F_1$ is the usual hypergeometric function. For the cases of interest
to us here the hypergeometric function can be simplified to yield:
\beq
J_{int}(3,y_m,y)=\frac{1}{8}[2y(y_m^2-y^2)^{3/2}+3y_m^2y(y_m^2y^2-y^2)^{1/2}
                 +3y_m^4\sin^{-1}(y/y_m)]
\label{psi.6}
\eeq
\barr
J_{int}(5,y_m,y) &= &\frac{1}{24}[4y(y_m^2-y^2)^{5/2}+5y_m^2y(y_m^2-y^2)^{3/2}\\
\nonumber
                & + &15y_m^4y(y_m^2-y^2)^{1/2} +15y_m^6\sin^{-1}(y/y_m)]
\label{psi.7}
\earr
\barr
J_{int}(7,y_m,y)&=&\frac{1}{192}[24y(y_m^2-y^2)^{7/2}+28y_m^2y(y_m^2-y^2)^{5/2}
                 +35y_m^4y(y_m^2)^{1/2}\\
\nonumber
               & + &105y_m^6y(y_m^2)^{1/2} +105y_m^8\sin^{-1}(y/y_m)]
\label{psi.8}
\earr
 The corresponding expressions for the Gaussian expressions for $\alpha_s$
cannot be done analytically. For the symmetric case, $\alpha_s=0$, one finds
that:
\beq
\Psi(x)=\frac{1}{2}[erf(x-1)-erf(x+1)]+2~erf(1)
\label{psi.9}
\eeq
with $erf(x)$ the error function:
\beq
erf(y)=\frac{2}{\pi}\int_0^y e^{-t^2}dt
\label{psi.10}
\eeq
The above functions $\Psi$ are plotted in Figs \ref{psi1} and  \ref{psi2}.

\begin{figure}
\includegraphics[height=0.3\textheight]{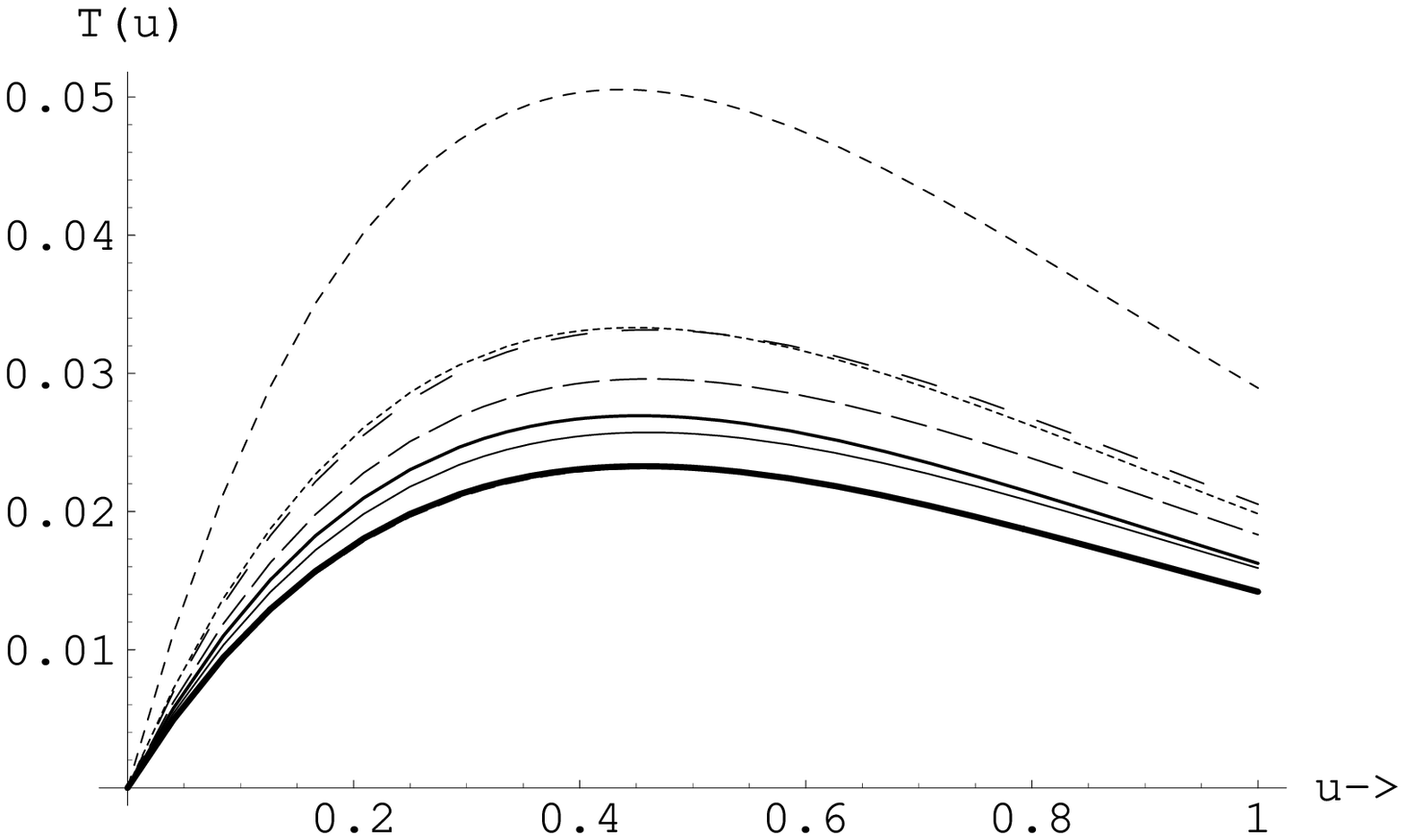}
\caption{The function T(u) for dark matter in the Eddington
theory for the coherent mechanism in the case of the target $^{127}I$. The
 graphs have been labeled as in Fig. \ref{psi2}. For illustration purposes 
an LSP mass of $100GeV$ has been chosen.
\label{T.u}
}
\end{figure}

From Fig. \ref{psi2} we see that both the value of the function $\Psi(x)$
as well as its range depend crucially on the parameters of the model. In the
absence of asymmetry the Gaussian model, with an appropriate escape velocity
put in by hand, yields results, which are almost four times larger than those
 of the Eddington
theory. In the presence of asymmetry our results in the Eddington theory
are substantially larger. This is due to the larger peak value attained as well
 as the larger allowed range of x. On the other hand we know that in the
Gaussian model the total rate is not significantly affected by the asymmetry
\citep{Verg99}.  The reason for the strong dependence of the results in the
 present model on the asymmetry is not,
of course, the asymmetry per se, but the fact that all parameters
change with it. In particular the value of $y_m$ (see table \ref{tab.1}).
In the Gaussian model the introduction of asymmetry did
 not affect the velocity distribution in any other way, e.g. it did not affect
the cutoff value of the velocity distribution.
The large upper values of $x$ in the function $\Psi(x)$, allowed in the present
model for large $y_m$, are, of course, somewhat
controlled by the nuclear form factor. In order to get a
 better feeling for such an effect on the rate, we also plot  the function
 $T(u)$, which is proportional
to the differential non directional rate. For illustration purposes we have
 chosen to present results for the popular target $^{127}I$ and a typical
 LSP mass of $100GeV$ (see Fig. \ref{T.u}). 
The results presented are for the coherent mechanism, but we expect very small
 changes when the spin contribution is considered.
It is clear that the introduction of asymmetry has a
profound effect on the rate. We have seen, however, that the large positive
values of the $\alpha_s$ can, in  principle,
 be eliminated from the data on the rotational curves.

 Effects like those discussed above
 may be more pronounced in the case of modulation, not studied in this
work.

\subsection{The directional rate}

The directional differential rate $(dR/du)_{dir}$ is proportional to
\barr
 T_{dir}(u)&=&\frac{1}{2\pi}~\bar{R}~a^2~|F(u)|^2~ \Psi_{dir}(a\sqrt{u})
\nonumber\\
       T_{dir}^{spin}&=&\frac{1}{2\pi}~\bar{R}_{spin}~a^2~|F_{11}u)| \Psi_{dir}(a\sqrt{u})
\label{T.3}
\earr
 with 
\beq
\Psi_{dir}(x)=~\int_0^x~ dy~\int_0^{\pi} ~ sin\theta d\theta
               \int_0^{2\pi}~d\phi~y~ f(y,\theta,\phi)~X~H(X)
\label{psi.11}
\eeq
with $H(X)$ the well known Heaviside function and $X$ is given by
\beq
X=\cos\Theta~\cos\theta+\sin\Theta~sin\theta[\sin\Phi~\sin\phi~+
~\cos\Phi~\cos\phi]
\label{X.1}
\eeq
where $\Theta$ and $\Phi$ describe the direction of observation $\hat{e}$
$$\hat{e}=\sin\Theta(\cos\Phi\hat{e}_x~+~\sin\Phi\hat{e}_y)~+~\cos\Theta\hat{e}_z$$
(There should be no confusion of the angle $\Phi$ used here with the potential
$\Phi(r)$ used earlier).
Note the presence of the factor of $1/(2\pi)$, since the azimuthal integration
of the recoiling nucleus is not present and we intend to use the same nucleon 
cross-section both in the directional and the non directional case.

The total rate is proportional to:
\beq
R_{dir}= ~\int_{u_{min}}^{u_{max}}(\frac{dR}{du})_{dir}du
\label{dir.1}
\eeq
Taking again the lowest Fourier components of the obtained rate as a function of
the phase of the Earth we get an expression similar to
Eq. \ref{tot.2}. Thus, to leading order in $\delta_1$, we can fit the total
 rate by an expression of the form:
\beq
R_{dir}= \frac{1}{2\pi}\bar{R}~t_{dir}~[1+(h_1-h_2)~\cos\alpha~+~h_3~\sin\alpha]
\label{dir.2}
\eeq
The parameters $t_{dir}$ and $h_i$ $i=1,2,3$ are obviously functions of the direction
 of observation, i.e. $\Theta$ and $\Phi$.
If one observes in the direction of the Sun's velocity $h_2=h_3=0$. Similarly
if one observes in a plane perpendicular to the Sun's velocity $h_1=0$. Instead of
$t_{dir}$ it is best to use the ratio: 
\beq
\kappa= 2\pi~\frac{R_{dir}}{R}=\frac{t_{dir}}{t}
\label{dir.3}
\eeq
 The parameter $\kappa$ is essentially independent of the LSP mass, the nuclear
 parameters and the asymmetry parameter $\alpha_s$. But it depends strongly
 on the direction of
 observation and is expected to correlate strongly with the angle between
$\hat{e}$ and the Sun's direction of motion. This correlation  provides a
an experimental signature perhaps
better the modulation with the Earth's motion in non directional experiments.

For $\delta_1 \neq 0$ the above integrals over $y,\theta,\phi$, especially in
the directional case,
can only be done numerically. Such results will appear elsewhere \citep{BOSV}.

\section{Conclusions}

In the present paper we studied the density and velocity distributions of
cold dark matter in the context of the Eddington theory, considering not only
symmetric but axially symmetric distributions as well. In our approach we used
standard simple distribution functions of the energy and angular momentum.
 This lead us to
simple relations between the density $\rho$ and the potential $\Phi$.
$\rho$ and $\Phi$ were obtained by solving numerically poisson's equation
 in a suitable region of space.
  This procedure allowed us to determine  the maximum permitted
 dark matter velocity.
 Then we were able to study both the rotational curves as
well as the velocity distribution of dark matter in our vicinity. We should
mention that this distribution is not Maxwellian and has an upper velocity
cutoff built into into it and not put in by hand as in the traditional
treatment with Gaussian distribution.

  Our results depend on a minimum set of
parameters, which were treated as free (see table \ref{tab.1}).

 We saw that in the context of this theory the predicted rotational
 velocities , see Figs \ref{f.vd1} - \ref{f.vs2}, depending on the input
parameters,
 can vary significantly in the presence of the asymmetry. This comes mainly
from the factor $x^2$ in the second term of Eq. \ref{dens.1b}, when
 the parameter $a$ is reasonably large.
 By comparing our results to the observed rotational curves
 \citep{Jungm}, one may constrain the parameters
of the model. The best
 choice seems to be the case with a
small asymmetry parameter $\alpha_s$.

 We have also made a preliminary study of the effect of the new velocity
 distribution on the direct detection rates for cold dark matter. For
 illustration purposes we have selected the case $\lambda=1/2$. We have seen
 that, in the context of
the Eddington approach, the total rates,
unlike the case of the Gaussian distribution \citep {Verg99,Verg00},
sensitively depend on the asymmetry parameter $\alpha_s$. This is due
to the fact that, when the asymmetry changes, the upper value of the
velocity distribution also changes.

It is thus not surprising that, in the Eddington theory,  the total
 (non directional and non modulated)
 event rates for direct LSP detection maybe 
substantially different from those of the phenomenological Gaussian
distributions (compare Figs \ref{psi1} and \ref{psi2}).
The strong dependence of the rate on the asymmetry parameter remains 
even after the nuclear form factor has been incorporated (see Fig. \ref{T.u}).
 Results of more detailed calculations of the event rates will appear elsewhere.

 The dependence of the directional and/or modulated rates on the velocity
distribution  is currently under study. We expect this dependence to be more
pronounced than on the total rates.

  \section{acknowledgments}
J.D.V. would like to thank the Physics Department of UNISA and Professor
S. Sofianos for their hospitality. D.O appreciates the hospitality provided 
by the University of Ioannina.

\end{document}